\documentclass[12pt]{article}
\usepackage[latin1]{inputenc}
\usepackage[twoside,left=1in,right=1in,top=1in,bottom=1in]{geometry}
\usepackage{amsmath}
\usepackage{amsfonts}
\usepackage{amssymb}
\usepackage{indentfirst}
\usepackage{fancyhdr}
\usepackage[dvips]{graphicx}
\usepackage{amssymb}
\usepackage{amsmath}
\usepackage{latexsym}
\usepackage{amsthm}
\usepackage{eucal}
\usepackage{natbib}
\usepackage[T1]{fontenc}
\usepackage{hyperref}
\usepackage{amsxtra}
\usepackage{amstext}
\usepackage{epsf,epsfig}
\usepackage{rotating}
\usepackage{multirow} 
\usepackage{pict2e}
\usepackage[table]{xcolor}
\usepackage{tikz}
\usetikzlibrary{shapes.geometric} 
\usepackage{morefloats}
\usepackage[bottom]{footmisc}

\usepackage{pifont}
\newcommand{\cmark}{\ding{51}}
\newcommand{\xmark}{\ding{55}}

\newcommand{\bzi}{\mathbf{z}_i} 
\newcommand{\bzj}{\mathbf{z}_j} 
\newcommand{\tbzio}{\tilde{\mathbf{z}}_{i}}
\newcommand{\tbzi}{\tilde{\mathbf{z}}_i}
\newcommand{\tbzj}{\tilde{\mathbf{z}}_j}

\newcommand{\bZ}{\mathbf{Z}} 
\newcommand{\bY}{\mathbf{Y}}

\newcommand{\tSigma}{\tilde{\boldsymbol{\Sigma}}}

\newcommand{\bI}{\mathbf{I}}
\newcommand{\yij}{y_{ij}}

\newcommand{\bbzi}{\boldsymbol{\bar{\mathbf{z}}}_i}
\newcommand{\bbzj}{\boldsymbol{\bar{\mathbf{z}}}_j}

\newcommand{\bbSigma}{\boldsymbol{\bar{\Sigma}}}
\newcommand{\txi}{\tilde{\xi}} 
\newcommand{\txio}{\tilde{\xi}_0} 
\newcommand{\txik}{\tilde{\xi}_k} 
\newcommand{\tpsi}{\tilde{\psi}}

\newcommand{\tSigmak}{\tilde{\boldsymbol{\Sigma}}_k}

\newcommand{\yijk}{y_{ijk}}

\newcommand{\bYa}{\mathbf{Y}_1}

\newcommand{\bYk}{\mathbf{Y}_k} 
\newcommand{\bYK}{\mathbf{Y}_K}

\newcommand{\ktbzi}{\tilde{\mathbf{z}}_{ik}} 
\newcommand{\tbziI}{\tilde{\mathbf{z}}_{i1}}
\newcommand{\ktbzj}{\tilde{\mathbf{z}}_{jk}}

\title{\bf Joint Modelling of Multiple Network Views}
\author{Isabella Gollini\thanks{Department of Civil Engineering,
University of Bristol, England.} \and  Thomas Brendan Murphy\thanks{School of Mathematical Sciences, Complex \& Adaptive Systems Laboratory and Insight Research Centre, University College Dublin, Ireland} }

\begin{document}
\maketitle
\begin{abstract}
Latent space models (LSM) for network data were introduced by \cite{HRH02} under the basic assumption that each node of the network has an unknown position in a $D$-dimensional Euclidean latent space: generally the smaller the distance between two nodes in the latent space, the greater their probability of being connected. 
In this paper we propose a variational inference approach to estimate the intractable posterior of the LSM.
In many cases, different network views on the same set of nodes are available. It can therefore be useful to build a model able to jointly summarise the information given by all the network views.
For this purpose, we introduce the latent space joint model (LSJM) that merges the information given by multiple network views assuming that the probability of a node being connected with other nodes in each network view is explained by a unique latent variable.
This model is demonstrated on the analysis of two datasets: an excerpt of 50 girls from `Teenage Friends and Lifestyle Study' data at three time points and the Saccharomyces cerevisiae genetic and physical protein-protein interactions.
\end{abstract}


\section{Introduction}\label{sec.intro}
Network data consists of a set of nodes and a list of edges between the nodes. Recently there has been a growing interest in the modelling of network data. A number of models have been proposed for network data including exponential random graph models (ERGMs) \citep{Holland1981}, stochastic blockmodels \citep{Holland1983,Air08} and latent space models \citep{HRH02,HRT07}. Recent reviews of various network modeling approaches include \cite{Goldenberg2010} and \cite{salter2012review}.

Latent space models (LSM) are a well known family of latent variable models for network data introduced by \cite{HRH02} under the basic assumption that each node has an unknown position in a $D$-dimensional Euclidean latent space: generally the smaller the distance between two nodes in the latent space, the greater the probability of them being connected.
Unfortunately, the posterior distribution of the LSM cannot be computed analytically. For this reason we propose a variational inferential approach which proves to be less computationally intensive than the MCMC procedure proposed in \cite{HRH02} and can therefore easily handle large networks.

In many cases, multiple network link relations on the same set of nodes are available.
Multiple network views, also known as multiplex networks \citep{Mucha10}, can be intended either as multiple link relations among the nodes of the network
or a single link relation observed over different conditions, such as one network evolving over time (longitudinal networks). 
In order to deal with multiplex networks we present a latent space joint model (LSJM) that merges the information given by the multiple network views by assuming that the probability of a node being connected with other nodes in each view is explained by a unique latent variable. 
To estimate this model we propose an EM algorithm: the parameter estimates obtained from fitting a LSM for each network view independently are used to approximate the joint posterior distribution of the LSJM; then these results are used to update the parameter estimates of every LSM. This process is iterated until convergence.
This model has a wide range of applications. For example in computer science it is of interest to summarize the different relations (e.g. friend, fan, follower or like) that we observe in social media sites like Facebook, Twitter and YouTube \citep{Tang11}. Another important application is in systems biology where the joint modeling of physical and genetic protein-protein interactions is of wide interest \citep{BKKI08}. Other contexts in which this model can be useful include social sciences and business marketing \citep{Ans11}.
The LSJM is demonstrated on the analysis of an excerpt of 50 girls from `Teenage Friends and Lifestyle Study' data at three time points \citep{PL00,PW03}, and two Saccharomyces cerevisiae networks \citep{Biogrid06}. 
All the methods of this paper are implemented in the \texttt{lvm4net} package for \texttt{R} \citep{R}. 

The paper is organized as follows. Section~\ref{sec.lsm} provides an introduction to latent space models for network data with a particular focus on the variational inference approach to fit the latent space model.
In Section~\ref{sec.lsjm} we introduce the latent space joint model for multiple network view data.
In Section~\ref{sec.miss} we show how missing link data can be managed using the LSJM.
In Section~\ref{sec.app} we illustrate the capabilities of the LSJM and we analyze its performance in the presence of missing edges by using cross-validation; the model is illustrated on the two example datasets (Section~\ref{sec.girls}-\ref{sec.pro}). 
We conclude, in Section~\ref{sec.concl} with a discussion of the model.

\section{Latent Space Model}\label{sec.lsm}
Latent space models for network data have been introduced by \cite{HRH02} under the basic assumption that each node $i$ has an unknown position $\bzi$ in a $D$-dimensional Euclidean latent space.
The {\it distance model} is an easy-to-interpret LSM which is based on the distance between the nodes in the latent space.
Generally the smaller the distance between two nodes in the latent space, the greater the probability that they connect.
This model supposes the network to be intrinsically symmetric since the distance between nodes in the latent space is symmetric and thus it has the feature of being reciprocal: if $\yij = 1$ then the probability of $y_{ji} = 1$ is large, where $\yij$ is the observed variable that is $1$ if we observe a link from node $i$ to node $j$, and $0$ otherwise. For this reason, the distance model is particularly suitable for undirected networks or directed networks that exhibit strong reciprocity.

Let $N$ is the number of observed nodes and let $\bY$ be the $N\times N$ adjacency matrix containing the network information, with entries $\yij$ (where $y_{ij}=0$ or $1$), and null diagonal. Let $\bZ$ is a $N \times D$ matrix of latent positions where each row is composed by $\bzi=(z_{i1},\ldots,z_{iD})$ the $D$-dimensional vector indicating the position of observation $i$ in the $D$-dimensional latent space.
The latent space model can be written as
\begin{equation*}
p(\bY|\bZ,\alpha)=\prod_{i\neq j}^Np(\yij|\bzi,\bzj,\alpha)=\prod_{i\neq j}^N \frac{\exp(\alpha-|\bzi-\bzj|^2)^{\yij}}{1+\exp(\alpha-|\bzi-\bzj|^2)}.
\end{equation*}
where for ease of notation $\prod_{i\neq j}^N$ is equivalent to $\prod_{i=1}^N\prod_{j=1,j\neq i}^N$.

\begin{figure}[!htp]
\begin{center}
\includegraphics[scale=.57]{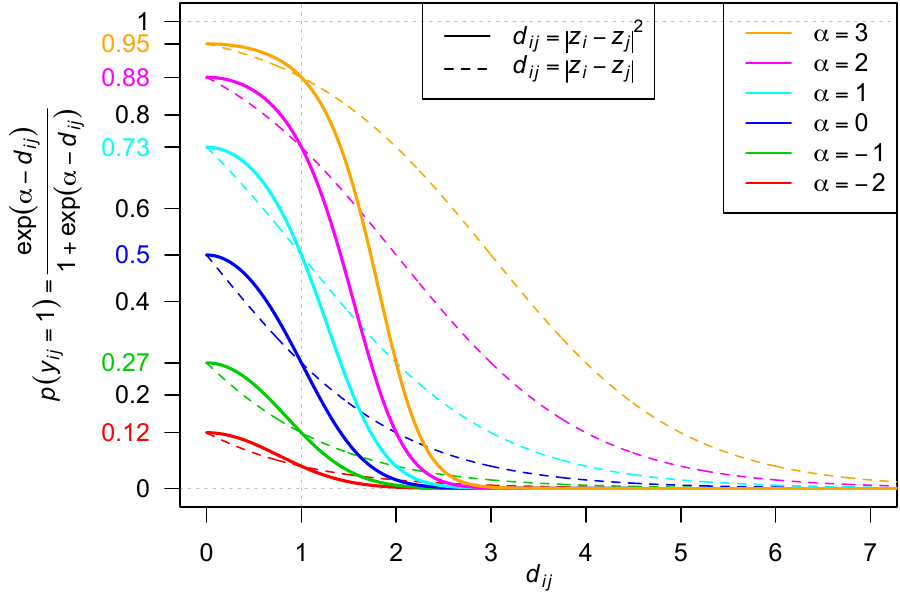}
\vspace{-.5cm}
\caption{Probability of $\yij=1$ as a function of the distance metrics. For $d_{ij} < 1$ the solid lines representing the probability of a link based on the squared Euclidean metric are higher then the dotted lines representing the  probability of a link based on the Euclidean distance. For $d_{ij}>1$ the solid lines decrease more rapidly then the dotted lines.}\label{fig:py1}
\end{center}
\end{figure}

We assume the following distributions for the model unknowns, where  $p(\alpha)=\mathcal{N}(\xi,\psi^2)$, $p(\bzi)\overset{iid}{=}\mathcal{N}(\mathbf{0},\sigma^2\mathbf{I}_D)$ and $\sigma^2,\xi,\psi^2$ are fixed parameters, and the squared Euclidean distance between observations $i$ and $j$ is $|\bzi-\bzj|^2=\sum_{d=1}^D(z_{id}-z_{jd})^2$.
The squared Euclidean distance measure is employed instead of the Euclidean distance used in \cite{HRH02}. This choice has been made for two main reasons: firstly, it allows one to visualize the data more clearly, giving an higher probability of a link between two close nodes in the latent space and lower probabilities to two nodes lying far away from each other (see Figure~\ref{fig:py1}); secondly it requires fewer approximation steps to be made in the estimation procedure.
An empirical comparison of the results obtained fitting a LSM using the Euclidean and the squared Euclidean distance is given in the supplementary material.

The posterior probability is of the unknown $(\bZ,\alpha)$ is of the form
\begin{equation*}
p(\bZ,\alpha|\bY)=Cp(\bY|\bZ,\alpha)p(\alpha)\prod_{i=1}^Np(\bzi),
\end{equation*}
where $C$ is the unknown normalizing constant.

\subsection{Variational Inference Approach}\label{sec.lsm.var}
Since the posterior distribution cannot be calculated analytically we make use of a  variational inference approach to estimate the model. To do this we aim  at maximizing a lower bound of the likelihood function. This approach has been proposed for several latent variable models \citep{Att99,JGJS99} and we refer to \cite{Bea03} for an extensive introduction to the variational methods. In the statistical network models context, \cite{Air08} proposed the use of the variational method to fit mixed-membership stochastic blockmodels and \cite{SalterTownshend2009} applied variational methods to fit the Latent Position Cluster Model \citep{HRT07}; the Latent Position Cluster Model is an extension of the original LSM in which the latent positions are assumed to come from a Gaussian mixture model.

We define the variational posterior $q(\bZ,\alpha|\bY)$ introducing the variational parameters $\Theta=(\txi,\tpsi^2)$, $\tbzi$ and $\tSigma$:
\begin{equation*}\label{q.var}
q(\bZ,\alpha|\bY)=q(\alpha)\prod_{i=1}^Nq(\bzi),
\end{equation*}
where $ q(\alpha)=\mathcal{N}(\txi,\tpsi^2)$ and $q(\bzi)=\mathcal{N}(\tbzi,\tSigma)$.

The basic idea behind the variational approach is to find a lower bound of the log marginal likelihood $\log p(\bY)$ by introducing the variational posterior distribution $q(\bZ,\alpha|\bY)$. 
This approach leads to minimize the Kulback-Leibler divergence between the variational posterior $q(\bZ,\alpha|\bY)$ and the true posterior $p(\bZ,\alpha|\bY)$:
\begin{equation*}\label{KL.var}
\begin{split}
\mathrm{KL}[q(\bZ,\alpha|\bY)||p(\bZ,\alpha|\bY)]&= -\int q(\bZ,\alpha|\bY) \log \frac{p(\bZ,\alpha|\bY)}{q(\bZ,\alpha|\bY)} \; d(\bZ,\alpha)\\
&=\int q(\bZ,\alpha|\bY) \log \frac{p(\bY,\bZ,\alpha)}{p(\bY) q(\bZ,\alpha|\bY)} \; d(\bZ,\alpha)\\
&=\int q(\bZ,\alpha|\bY) \log \frac{p(\bY,\bZ,\alpha)}{q(\bZ,\alpha|\bY)} \; d(\bZ,\alpha) - \int q(\bZ,\alpha|\bY) \log p(\bY) \; d(\bZ,\alpha)\\
&=\int q(\bZ,\alpha|\bY) \log \frac{p(\bY,\bZ,\alpha)}{q(\bZ,\alpha|\bY)} \; d(\bZ,\alpha)  - \log p(\bY) 
\end{split}
\end{equation*}
The last line follows as $\log p(\bY)$ is neither a function of $\bZ$ and $\alpha$. 
From this equation it is evident that minimizing $\mathrm{KL}[q(\bZ,\alpha|\bY)||p(\bZ,\alpha|\bY)]$ corresponds to maximizing the following lower bound:
\begin{equation*}
 \log p(\bY)\geq \int q(\bZ,\alpha|\bY) \log \frac{p(\bY,\bZ,\alpha)}{q(\bZ,\alpha|\bY)} \; d(\bZ,\alpha)
\end{equation*}
The Kulback-Leibler divergence between the variational posterior and the true posterior for the LSM can be written as:
\begin{equation*}\label{KL.var}
\begin{split}
\mathrm{KL}[q(\bZ,\alpha|\bY)||p(\bZ,\alpha|\bY)]&=\mathrm{KL}[q(\alpha) || p(\alpha)] + \sum_{i=1}^N \mathrm{KL}[q(\bzi) || p(\bzi)] - \mathbb{E}_{q(\bZ,\alpha|\bY)}[\log(p(\bY|\bZ,\alpha))\\
&= \frac{1}{2}\left(\frac{\tpsi^2}{\psi^2} - \log \frac{\tpsi^2}{\psi^2} + \frac{(\txi-\xi)^2}{\psi^2} +ND\log(\sigma^2) - N \log ( \det (\tSigma)) \right)\\
&\quad +  \frac{N}{2\sigma^2}\mathrm{tr}(\tSigma) + \frac{\sum_{i=1}^N\tbzi^T\tbzi}{2\sigma^2} - \mathbb{E}_{q(\bZ,\alpha|\bY)}[\log(p(\bY|\bZ,\alpha))] - \frac{1+ND}{2},
\end{split}
\end{equation*}
where the expected log-likelihood $\mathbb{E}_{q(\bZ,\alpha|\bY)}[\log(p(\bY|\bZ,\alpha))]$ is approximated using the Jensen's inequality:
\begin{equation}\label{ll.var}
\begin{split}
\mathbb{E}_{q(\bZ,\alpha|\bY)}[\log(p(\bY|\bZ,\alpha))]&=\sum_{i\neq j}^N\yij\mathbb{E}_{q(\bZ,\alpha|\bY)}[\alpha-|\bzi-\bzj|^2]-\mathbb{E}_{q(\bZ,\alpha|\bY)}[\log(1+\exp(\alpha-|\bzi-\bzj|^2))] \\
&\leq \sum_{i\neq j}^N\yij(\mathbb{E}_{q(\bZ,\alpha|\bY)}[\alpha -|\bzi-\bzj|^2])-\log(1+\mathbb{E}_{q(\bZ,\alpha|\bY)}[\exp(\alpha-|\bzi-\bzj|^2)])\\
&= \sum_{i\neq j}^N\yij(\txi-2 \mathrm{tr}(\tSigma)-|\tbzi-\tbzj|^2)\\
&\quad -\log \left(1+\frac{\exp\left(\txi+\dfrac{1}{2}\tpsi^2\right)}{\det(\bI+4\tSigma)^{\frac{1}{2}}}\exp\left(-(\tbzi-\tbzj)^T(\bI+4\tSigma)^{-1}(\tbzi-\tbzj)\right)\right).
\end{split}
\end{equation}

From this equation it is possible to understand the computational advantage of the squared Euclidean distance model with respect to the Euclidean distance model. In fact, the expected log-likelihood has been approximated using the Jensen's inequality whereas \cite{SalterTownshend2009} need to use three first-order Taylor-expansions to fit the model with the Euclidean distance.
An alternative approach to approximate the expected log-likelihood is given by the Jaakola \& Jordan bound \citep{JJ00}, but it would require further approximations, and it would be more difficult to compute.

To estimate the model an EM algorithm \citep{Dempster77} can be applied. The EM algorithm consists of two main steps: the first step, called the E-step, aims to estimate the parameters $\tbzi,\tSigma$ of the posterior distribution of the latent space positions by maximizing the complete data log-likelihood given all the other parameters $\Theta$. The second step is the M-step where $\Theta$ is updated maximizing  the complete data log-likelihood given $\tbzi$ and $\tSigma$. As observed above, in this context maximizing the log-likelihood corresponds to minimize the Kullback-Leibler divergence between the variational posterior and the true posterior. Therefore the EM algorithm can be written as a function of the Kullback-Leibler divergence, this approach is commonly known as Variational EM algorithm \citep{JGJS99}. 
This method scales as $\mathcal{O}(N^2)$, but it converges in just a few iterations, and the calculations performed in the estimation procedure are pretty simple (see the supplementary material for a comparison of CPU times with other methods and models). 

The analytical form of the parameter estimates will be found introducing the first and second order Taylor series expansion approximation of the following function:
\begin{equation}\label{lsjm.lsm.f}
f(\tbzi,\tSigma,\txi,\tpsi^2)=
\log \left(1+\frac{\exp\left(\txi+\dfrac{1}{2}\tpsi^2\right)}{\det(\bI+4\tSigma)^{\frac{1}{2}}}\exp\left(-(\tbzi-\tbzj)^T(\bI+4\tSigma)^{-1}(\tbzi-\tbzj)\right)\right)
\end{equation}
calculated around the estimates calculated at the previous step of the algorithm (see supplementary material).

Here we outline the Variational EM algorithm on the $(i + 1)th$ iteration:\\
\noindent {\bf E-Step} Estimate the parameters of the latent posterior distributions $\tbzi^{(i+1)}$ and $\tSigma^{(i+1)}$ evaluating:
\begin{equation*}
\begin{split}
\mathcal{Q}(\Theta;\Theta^{(i)})&=-\mathrm{KL}(q(\bZ,\alpha|\bY)||p(\bZ,\alpha|\bY))\\
&=\int q(\bZ,\alpha|\bY) \log \frac{p(\bZ,\alpha|\bY)}{q(\bZ,\alpha|\bY)} \; d(\bZ,\alpha),
\end{split}
\end{equation*}
where $\Theta=(\txi,\tpsi^2)$. This gives
\begin{equation*}
\tSigma^{(i+1)}=\frac{N}{2}\left[ \left(\frac{N}{2\sigma^2}+2\sum_{i=1}^N\sum_{j\neq i}\yij\right)\bI +J(\tSigma^{(i)}) \right]^{-1},
\end{equation*}
where $J$ is the Jacobian matrix of $f(\tbzi^{(i)},\tSigma^{(i)},\txi^{(i)},\tpsi^{2(i)})$ (Equation \ref{lsjm.lsm.f}) evaluated at $\tSigma=\tSigma^{(i)}$. And
\begin{equation}\label{eq:z.lsm}
\tbzi^{(i+1)}=\left[ \left(\frac{1}{2\sigma^2}+\sum_{j \neq i}(y_{ji}+y_{ij})\right)\bI + H(\tbzi^{(i)}) \right]^{-1}\left[\sum_{j \neq i}(y_{ji}+y_{ij})\tbzj-G(\tbzi^{(i)})+H(\tbzi^{(i)}) \tbzi^{(i)} \right],
\end{equation}
where $G$ is the gradient and $H$ is the Hessian matrix of $f(\tbzi^{(i)},\tSigma^{(i+1)},\txi^{(i)},\tpsi^{2(i)})$ evaluated at $\tbzi=\tbzi^{(i)}$.\\
\noindent {\bf M-Step} Estimate the parameters of the posterior distribution of $\alpha$ evaluating:
\begin{equation*}
\Theta^{(i+1)}= \mathrm{argmax} \; \mathcal{Q}(\Theta;\Theta^{(i)}).
\end{equation*}

This gives
\begin{equation*}
\txi^{(i+1)}=\frac{\xi+\psi^2(\sum_{i=1}^N\sum_{j\neq i} \yij-f'(\txi^{(i)})+\txi^{(i)} f''(\txi^{(i)}))}{1+\psi^2 f''(\txi^{(i)})},
\end{equation*}
where $f'$ and $f''$ are the first and the second derivatives of $f(\tbzi^{(i+1)},\tSigma^{(i+1)},\txi^{(i)},\tpsi^{2(i)})$ evaluated at $\txi=\txi^{(i)}$, and
\begin{equation*}
\tpsi^{2(i+1)}=\left(\frac{1}{\psi^2} +2 f'(\tpsi^{2(i)}) \right)^{-1},
\end{equation*}
where $f'$ is the first derivative of $f(\tbzi^{(i+1)},\tSigma^{(i+1)},\txi^{(i+1)},\tpsi^{2(i)})$ evaluated at $\tpsi^2=\tpsi^{(i)2}$.

\section{Latent Space Joint Model}\label{sec.lsjm}

\begin{figure}[!htp]
\begin{center}
\includegraphics[scale=1]{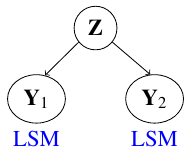}
\vspace{-.5cm}
\caption{Latent space joint model for two network views.}\label{fig:zy1y2}
\end{center}
\end{figure}

Let us suppose to have $K$ network views on the same set of $N$ nodes. We introduce a model which assumes that a continuous latent variable $\bzi\sim \mathcal{N}(0,\sigma^2\mathbf{I}_D)$ is able to summarize the information given by all the network views $\bYa,\ldots,\bYK$ identifying the position of node $i$ in a $D$-dimensional latent space. In this case the network data assumes conditionally independence given the latent variable.
Our purpose is to model each network view by using a LSM (see Figure~\ref{fig:zy1y2}). 

This yields to the following joint model:
\begin{equation*}
\begin{split}
p(\bYa,\ldots,\bYK|\bZ,\alpha_1,\ldots,\alpha_K)&=\prod_{k=1}^K p(\bYk|\bZ;\alpha_{k})\\
&=\prod_{k=1}^K\prod_{i\neq j}^N \frac{\exp(\alpha_k-|\bzi-\bzj|^2)^{\yijk}}{1+\exp(\alpha_k-|\bzi-\bzj|^2)}
\end{split}
\end{equation*}
where $p(\alpha_k)=\mathcal{N}(\xi_k,\psi_k^2)$, $p(\bzi)\overset{iid}{=}\mathcal{N}(\mathbf{0},\sigma^2\mathbf{I}_D)$ with $\sigma^2,\xi_k,\psi_k^2$ set to be fixed parameters, and the dyad $\yijk$ takes value $1$ if there is a link between node $i$ and node $j$ in network $k$, and $0$ otherwise.

The following identity allows one to find the model parameters $\alpha_{1},\ldots,\alpha_{K}$ and the posterior distribution of the latent variable $\bzi$ given the $K$ models:
\begin{equation}
\begin{split}
p(\bzi|\bYa,\ldots,\bYK;\alpha_{1},\ldots,\alpha_{K})&\propto p(\bzi) \prod_{k=1}^K p(\bYk|\bzi;\alpha_{k})\\
&\propto p(\bzi)\prod_{k=1}^K  \dfrac{p(\bzi|\bYk;\alpha_{k}) }{p(\bzi)}\\
&=\dfrac{ \prod_{k=1}^K p(\bzi|\bYk;\alpha_{k})}{p(\bzi)^{K-1}}.
\end{split}
\end{equation}\label{eq:pzyy}
Applying the variational inference approach presented in Section~\ref{sec.lsm.var} introducing the variational parameters $\Theta_k=(\txi_k,\tpsi_k^2)$, $\ktbzi$ and $\tSigmak$, it is possible to approximate $p(\bzi|\bYk;\alpha_{k})$ with $q(\bzi) \sim \mathcal{N}(\ktbzi,\tSigmak)$. 
Recalling that $\bzi\sim \mathcal{N}(\mathbf{0},\sigma^2\mathbf{I}_D)$ and Equation~\ref{eq:pzyy}, we obtain that the posterior distribution of the latent variables given all the network views can be written as:
\begin{equation*}
p(\bzi|\bYa,\ldots,\bYK;\Theta_{1},\ldots,\Theta_{K})\propto \mathcal{N}(\bbzi,\bbSigma),
\end{equation*}
where the parameters are 
\begin{equation} \label{eq:zb.lsjm}
\bbSigma=\left[ \sum_{k=1}^K \tSigmak^{-1} -\dfrac{K-1}{\sigma^2}\mathbf{I}_D\right]^{-1}
\qquad
\mathrm{and }
\qquad
\bbzi
=\bbSigma\left[\sum_{k=1}^K \tSigmak^{-1} \ktbzi  \right].
\end{equation}
By fitting LSJM we get information on $p(\bzi|\bYa,\ldots,\bYK;\Theta_{1},\ldots,\Theta_{K})$ and $p(\bzi|\bYk;\Theta_{k})$ for $k=1,\ldots,K$. This way it is possible to have estimates for both the overall positions $\bbzi$ and the position given one particular network view $\ktbzi$.

The estimates of $\ktbzi$ and $\tSigmak$ are updated from $\bbzi$ and $\bbSigma$, so we can locate the unconnected nodes or subgraphs in the latent space depending on their position conditional on the other network views, avoiding the usual tendency of pushing away the unconnected nodes to maximize the likelihood when fitting the classical LSM. This approach also allows one to plot the positions given each network view in the same latent space, and to look at how the nodes in each network change the positions.

Let $(\Theta^{(i)}_{1},\ldots,\Theta^{(i)}_{K})$ be the current estimates of $(\Theta_{1},\ldots,\Theta_{K})$ and initialize $(\Theta^{(0)}_{1},\ldots,\Theta^{(0)}_{K})$.
The Variational EM algorithm at the iteration $(i+1)$ can be summarized as follows:

\noindent {\bf E-Step} Estimate the parameters $\bbSigma^{(i+1)}$ and $\bbzi^{(i+1)}$ of the posterior distribution of the latent variables given all the network views evaluating:
\begin{equation}\label{eq.ell}
\begin{split}
\mathcal{Q}(\Theta_{1},&\ldots,\Theta_{K};\Theta^{(i)}_{1},\ldots,\Theta^{(i)}_{K})=\\
&=\mathbb{E}_{p(\bZ|\bYa,\ldots,\bYK;\Theta^{(i)}_{1},\ldots,\Theta^{(i)}_{K}) }[\log(p(\bYa,\ldots,\bYK,\bZ|\Theta_{1},\ldots,\Theta_{K}))] \\
&=\sum_{k=1}^K\mathbb{E}_{p(\bZ|\bYa,\ldots,\bYK;\Theta^{(i)}_{1},\ldots,\Theta^{(i)}_{K}) }[\log(p(\bYk,\bZ|\Theta_{k}))]\\
&\quad -(K-1)\mathbb{E}_{p(\bZ|\bYa,\ldots,\bYK;\Theta^{(i)}_{1},\ldots,\Theta^{(i)}_{K}) }[\log(p(\bZ))].
\end{split}
\end{equation}

Thus we can estimate the parameters of the posterior distribution $p(\bzi|\bYk;\Theta_{k})$  given each network separately:
\begin{equation*}
\tSigmak^{(i+1)}=\frac{N}{2}\left[ \left(\frac{N}{2\sigma^2}+2\sum_{i=1}^N\sum_{j\neq i}\yijk\right)\bI +J_k(\bbSigma^{(i)}) \right]^{-1},
\end{equation*}
where $J_k$ is the Jacobian matrix of $f(\bbzi^{(i)},\bbSigma^{(i)},\txi_k^{(i)},\tpsi_k^{2(i)})$ (Equation~\ref{lsjm.lsm.f}) evaluated at $\tSigma=\bbSigma^{(i)}$, and,
\begin{equation}\label{eq:z.lsjm}
\ktbzi^{(i+1)}=\left[ \left(\frac{1}{2\sigma^2}+\sum_{j\neq i}(y_{jik}+y_{ijk})\right)\bI + H_k(\bbzi^{(i)}) \right]^{-1}\left[\sum_{j\neq i}(y_{jik}+y_{ijk})\bbzj^{(i)}-G_k(\bbzi^{(i)})+H_k(\bbzi^{(i)}) \bbzi^{(i)} \right],
\end{equation}
where $G_k$ and $H_k$ are respectively the gradient and the Hessian matrices of $f(\bbzi^{(i)},\bbSigma^{(i+1)},\txi_k^{(i)},\tpsi_k^{2(i)})$ evaluated at $\tbzi=\bbzi^{(i)}$.

The posterior distribution of the latent positions given all the network views is estimated merging the estimates of the single models:
\begin{equation*}
p(\bzi|\bYa,\ldots,\bYK;\Theta_{1},\ldots,\Theta_{K}) \propto \mathcal{N}(\bbzi,\bbSigma),
\end{equation*}
where
\begin{equation*}
\bbSigma^{(i+1)}=\left[ \sum_{k=1}^K [\tSigmak^{(i+1)}]^{-1} -\dfrac{K-1}{\sigma^2}\mathbf{I}_D\right]^{-1},
\qquad
\mathrm{and }
\qquad
\bbzi^{(i+1)}=\bbSigma^{(i+1)}\left[\sum_{k=1}^K [\tSigmak^{(i+1)}]^{-1} \ktbzi^{(i+1)} \right].
\end{equation*}\\
\noindent {\bf M-Step} Update the model parameters evaluating
\begin{equation*}
(\Theta^{(i+1)}_{1},\ldots,\Theta^{(i+1)}_{K})= \mathrm{argmax} \; \mathcal{Q}(\Theta_{1},\ldots,\Theta_{K};\Theta^{(i)}_{1},\ldots,\Theta^{(i)}_{K}).
\end{equation*}
This gives
\begin{equation*}
\txi^{(i+1)}_k=\frac{\xi_k+\psi_k^{2}(\sum_{i=1}^N\sum_{j\neq i} \yijk-f_k'(\txik^{(i)})+\txik^{(i)}f_k''(\txik^{(i)}))}{1+\psi_k^{2} f_k''(\txik^{(i)})},
\end{equation*}
where $f_k'$ and $f_k''$ are the first and the second derivatives of $f(\bbzi^{(i+1)},\bbSigma^{(i+1)},\txi_k^{(i)},\tpsi_k^{2(i)})$ evaluated at $\txi=\txik^{(i)}$, and
\begin{equation*}
\tpsi_k^{2(i+1)}=\left(\frac{1}{\psi_k^2} +2 f_k'(\tpsi_k^{(i)2}) \right)^{-1},
\end{equation*}
where $f_k'$ is the first derivative of $f(\bbzi^{(i+1)},\bbSigma^{(i+1)},\txi_k^{(i+1)},\tpsi_k^{2(i)})$ evaluated at $\tpsi^2=\tpsi_k^{(i)2}$.

\section{Missing Link Data}\label{sec.miss}
Recent Bayesian approaches to predict missing links in network data have been introduced by \cite{Hoff09} who proposed to use multiplicative latent factor models, and \cite{Koskinen10,Koskinen13} in the context of Bayesian exponential random graph models. However there is lack of methods able to take into account the information given by the presence of multiple network views.

Missing (unobserved) links can be easily managed by the LSJM using the information given by all the network views. To estimate the probability of the presence ($y_{ijk}=1$) or absence ($y_{ijk}=0$) of an edge we employ the posterior mean of the $\alpha_k$ and of the latent positions so that we get the following equation:
\begin{equation*}
\yijk^*=p(\yijk=1|\bbzi,\bbzj,\txik)=\frac{\exp(\txik-|\bbzi-\bbzj|^2)}{1+\exp(\txik-|\bbzi-\bbzj|^2)}.
\end{equation*}
If we want to infer whether to assign $\yijk=1$ or $\yijk=0$, we need to introduce a threshold $\tau_k$, and let $\yijk=1$  if $p(\yijk=1|\ktbzi,\ktbzj,\txik)> \tau_k$, and $\yijk=0$ otherwise. We set the threshold to be equal to the median probability of a link for the subset of the actual observed links in network $k$.

To evaluate link prediction in presence of missing links, we use a 10-fold cross validation procedure consisting of randomly splitting the set of all the possible dyads  in each network view into 10 subsets. We can predict the links of each subset given the others fitting a LSM to each network independently and then fitting the LSJM.  Finally we can compare the link prediction performance given by these two methods.

The LSJM allows one to locate in the latent space a missing node (no information about links sent and received by the node) in one network by employing the information provided by the other network views. We evaluate the link prediction approach for the missing nodes applying a 10-fold cross validation on the nodes, randomly dividing the set of nodes in each network into 10 subsets, and then predicting the links using the LSJM. We do not use a single LSM to locate missing nodes in the latent space and to estimate their probabilities of links since the only information that the model would use estimating of the latent positions would be the prior distribution of the nodes $p(\bzi)$.

To facilitate the interpretation of the results we matched the rotation of the latent positions in the single LSM with the ones obtained from the LSJM.

\section{Applications}\label{sec.app}

\subsection{Computational Aspects}\label{sec.app.comp}

The LSM and the LSJM have been fitted assuming that $p(\alpha)=\mathcal{N}(0,2)$ and $p(\bzi)\overset{iid}{=}\mathcal{N}(\mathbf{0},\mathbf{I}_2)$, initializing the variational parameters $\txik=0$ and $\tpsi^2_k=2$, and latent positions $\bbzi$ by random generated numbers from $\mathcal{N}(\mathbf{0},\mathbf{I}_2)$ and setting $\bbSigma=\mathbf{I}_2$.

We have set the latent space to be bi-dimensional in order to be able to visualize and easily interpret the results.
Ten random starts of the algorithm were used and the solution returning the maximum expected likelihood value was selected.
The latent positions are identifiable up to a rotation of the latent space. For this reason to speed up the convergence of the algorithm we matched the estimates of $\ktbzi$ $\forall k>1$ with $\tbziI$ via singular value decomposition in the first 10 iterations of the EM algorithm.

The EM algorithm was stopped after at least 10 iterations when
\begin{equation*} \label{emstop}
\| \mathbb{E}_q[\log(p(\bY|\bZ,\alpha))]^{(i+1)}-\mathbb{E}_q[\log(p(\bY|\bZ,\alpha))]^{(i)}\|<{tol},
\end{equation*}
where $\mathbb{E}_q[\log(p(\bY|\bZ,\alpha))]$ is given by Equation~\ref{ll.var} if we fit the LSM or by Equation~\ref{eq.ell} if we fit the LSJM, $i$ indicates the iteration, ${tol}$ is a desired tolerance value (which, in this case, was set ${tol}=10^{-2}$).

To assess the fit of the model we evaluated the in-sample predictions producing the ROC curve of the estimated link probabilities and calculating the area under the curve (AUC) and the boxplots of the estimated link probabilities for both the true positive and the true negative links. We estimated the probability of a link under each network view by calculating:
\begin{equation*}
p(\yijk=1|\ktbzi,\ktbzj,\txik)=\frac{\exp(\txik-|\ktbzi-\ktbzj|^2)}{1+\exp(\txik-|\ktbzi-\ktbzj|^2)}.
\end{equation*}
All the calculations have been done using the \texttt{R} package \texttt{lvm4net}. 

\subsection{Excerpt of 50 girls from `Teenage Friends and Lifestyle Study'}
\label{sec.girls}

\begin{figure}[!htp]
\begin{center}
\includegraphics[scale=.8]{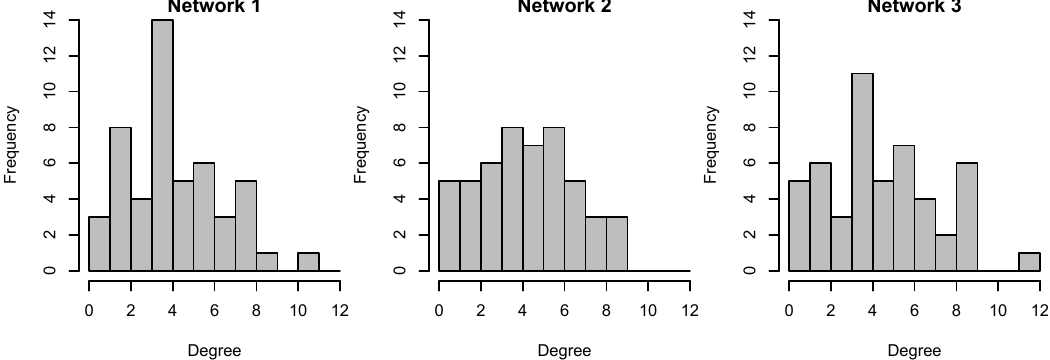}
\vspace{-.5cm}
\caption{Degree Distribution for the Girls Networks.}\label{fig:degrees50}
\end{center}
\end{figure}

\cite{PL00} and \cite{PW03} collected data for a `Teenage Friends and Lifestyle Study'. The dataset contains three directed networks about friendship relations between students in a school in Glasgow, Scotland. Each student was asked to name up to six best friends in the cohort. The data comes from three yearly waves, from 1995 to 1997. 
An extended description of all the data in the study can be found in \cite{PL00}.   In this paper we will focus on an excerpt of 50 girls that were present at all three measurement points. This dataset is available at the SIENA software website\footnote{\url{http://www.stats.ox.ac.uk/~snijders/siena/}} (\cite{RSP11}).
The networks are formed of 113, 116 and 122 links respectively, and have density of 0.046, 0.047, 0.049. Their degree distributions are shown in Figure~\ref{fig:degrees50}.

\subsubsection{LSM}
\label{sec.lsm.app.girls}

In this section we applied the LSM to the excerpt of 50 girls from the `Teenage Friends and Lifestyle Study' data. 

\begin{figure}[!htp]
\begin{center}
\includegraphics[scale=.47]{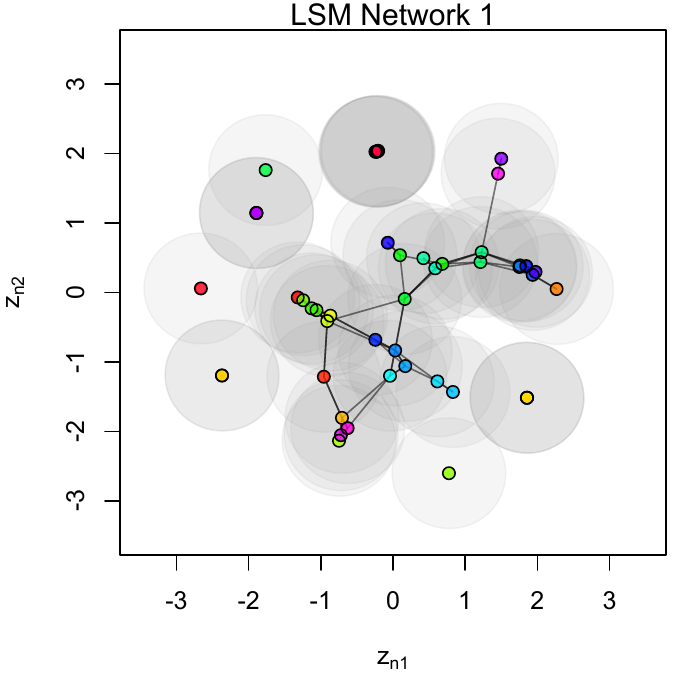}
\includegraphics[scale=.47]{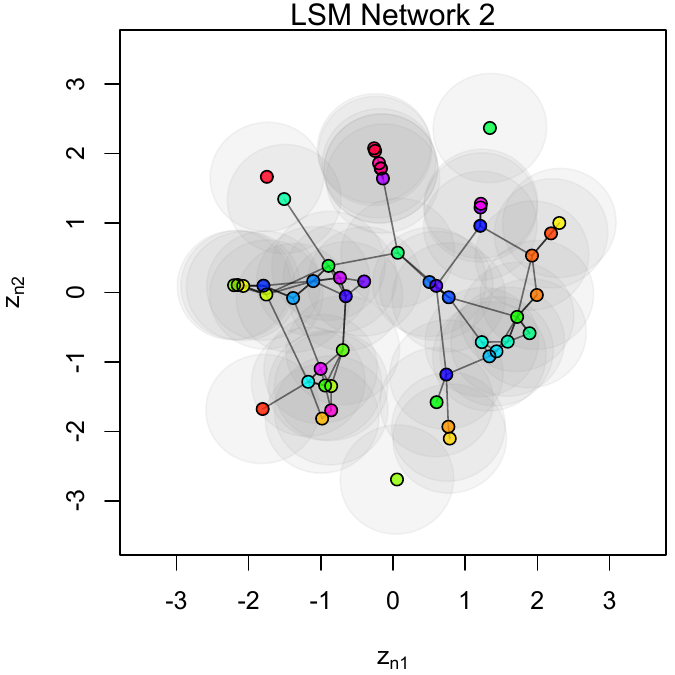}
\includegraphics[scale=.47]{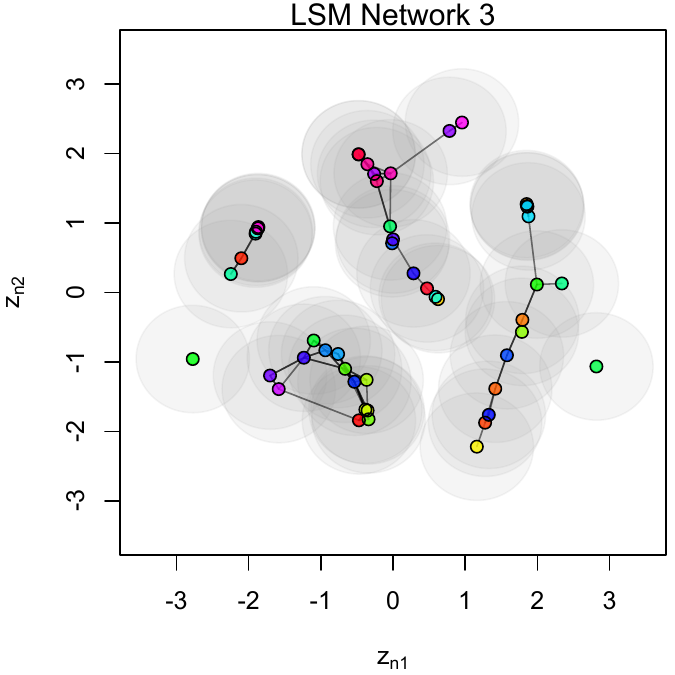}
\vspace{-.5cm}
\caption{Latent positions $\tbzi$ for the Girls Networks fitting the LSM. The grey ellipses represent the $95 \%$ approximate credible intervals. Overlapping approximate credible intervals make darker shades of grey.}\label{fig:girls.lsm}
\end{center}
\end{figure}

\begin{figure}[htp]
\begin{center}
\includegraphics[scale=0.5]{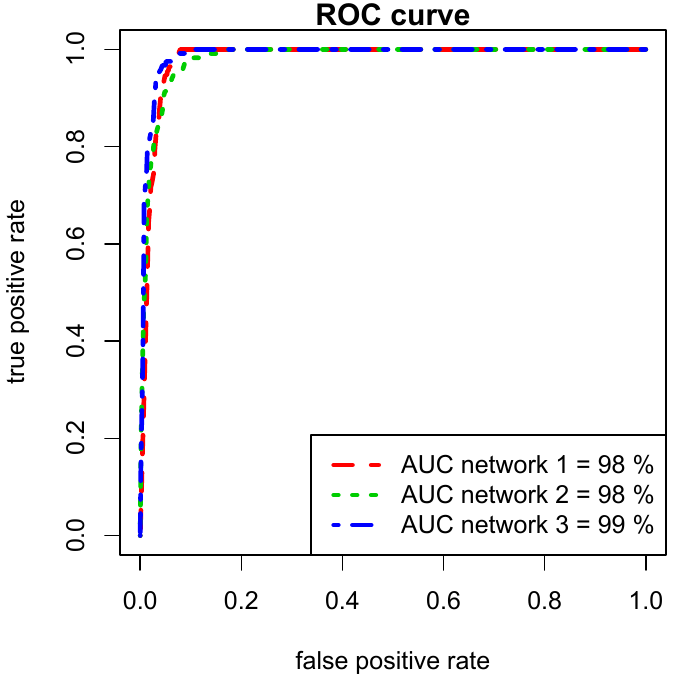}\hspace{.5cm}
\includegraphics[scale=0.5]{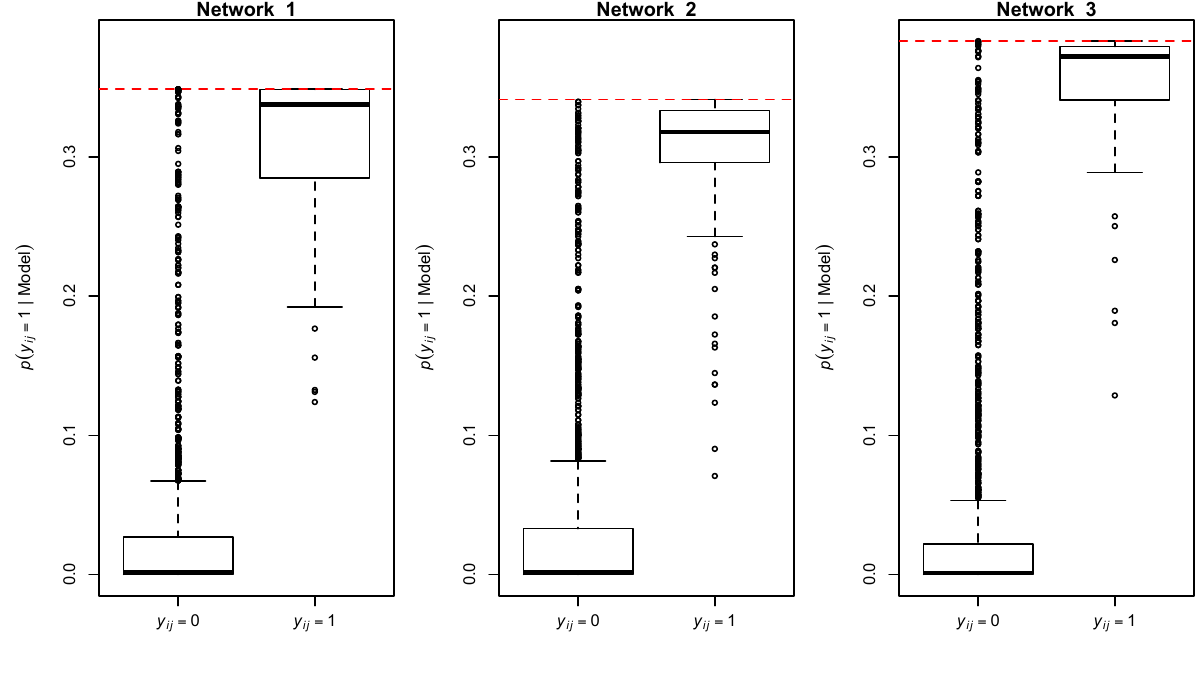}
\caption{ROC curves (left) and boxplots (right) of the estimated probabilities of a link for the true negatives and true positives obtained by fitting the LSM to the three Girls Networks}\label{fig:girls.lsm.roc}
\end{center}
\end{figure}

We fitted a LSM to each network separately assuming that $p(\alpha)=\mathcal{N}(0,2)$ and $p(\bzi)\overset{iid}{=}\mathcal{N}(\mathbf{0},\mathbf{I}_2)$. The estimated latent posterior positions $\tbzi$ (defined in Equation~\ref{eq:z.lsm}) are shown in Figure~\ref{fig:girls.lsm}. A matching rotation of the final estimates was applied in order to facilitate the interpretation of the results. The posterior distributions for the $\alpha$ parameters are quite similar over time $q(\alpha_1)=\mathcal{N}(-0.63,0.01)$, $q(\alpha_2)=\mathcal{N}(-0.66,0.01)$ and $q(\alpha_3)=\mathcal{N}(-0.48,0.01)$; this means that there are not big changes in terms of network density over time.
From the analysis of the ROC curve (Figure~\ref{fig:girls.lsm.roc} left) and AUC of the in-sample estimated link probabilities, it seems clear that the proposed LSM fits quite well the data separately. The boxplots in Figure~\ref{fig:girls.lsm.roc} (right) show that the estimated probabilities distinguish quite well the true negatives with low probability of forming a link from the true positives with high probability of forming a link.

\subsubsection{LSJM}\label{sec.app.girls}
\begin{figure}[!htp]
\begin{center}
\includegraphics[scale=.47]{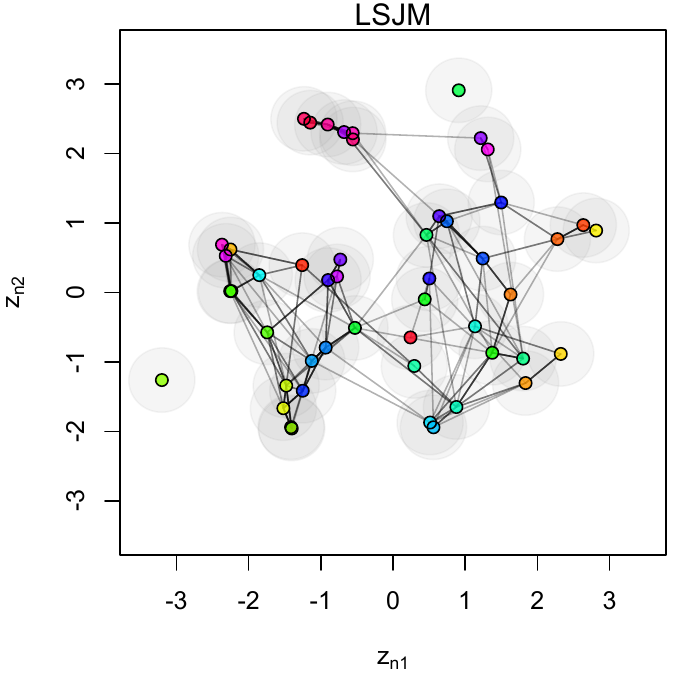}
\includegraphics[scale=.47]{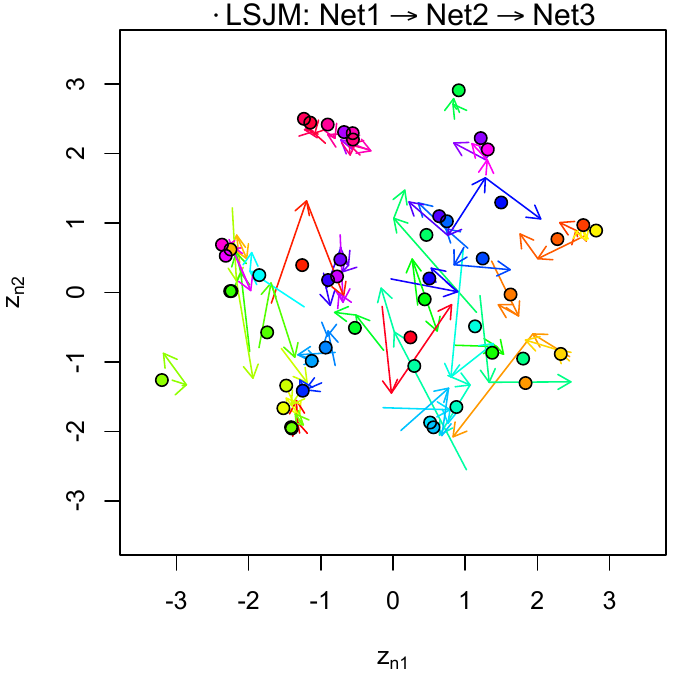}
\vspace{-.5cm}
\caption{On the left are shown the latent positions $\bbzi$ for the Girls Networks fitting the LSJM. 
In the plot on the right the dots represent the overall positions $\bbzi$ and the arrows connect the estimated position under each model $\tilde{\mathbf{z}}_{i1},\tilde{\mathbf{z}}_{i2},\tilde{\mathbf{z}}_{i3}$.}\label{fig:girls.lsjm}
\end{center}
\end{figure}

In this section we fitted the LSJM to the Girls dataset. The overall positions $\bbzi$ (defined in Equation~\ref{eq:zb.lsjm}) in the latent space are displayed in Figure~\ref{fig:girls.lsjm}. In Figure~\ref{fig:girls.lsjm} (right) the dots represent the overall positions $\bbzi$ and the arrows connect the estimated positions $\ktbzi$ under each model $k$ (defined in Equation~\ref{eq:z.lsjm}) from network $k=1$ to $k=3$, so that it is possible to see their variation over time.

\begin{figure}[!htp]
\begin{center}
\includegraphics[scale=.47]{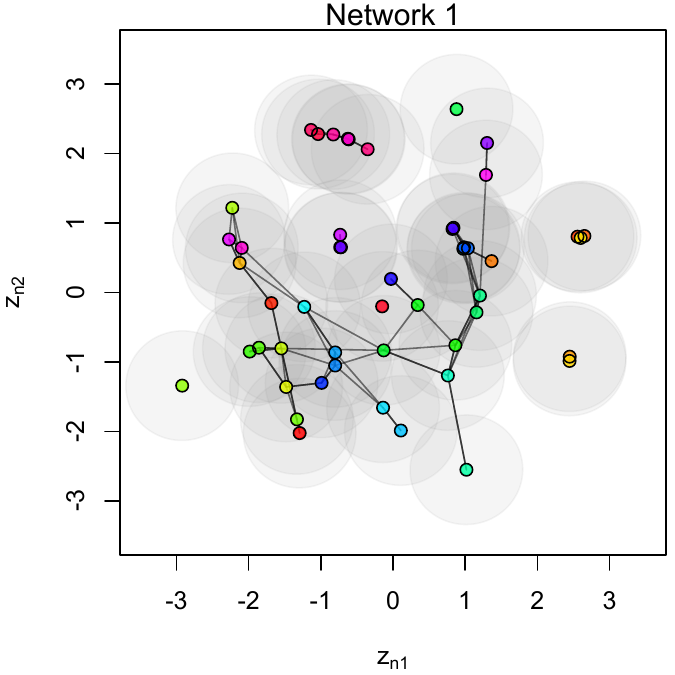}
\includegraphics[scale=.47]{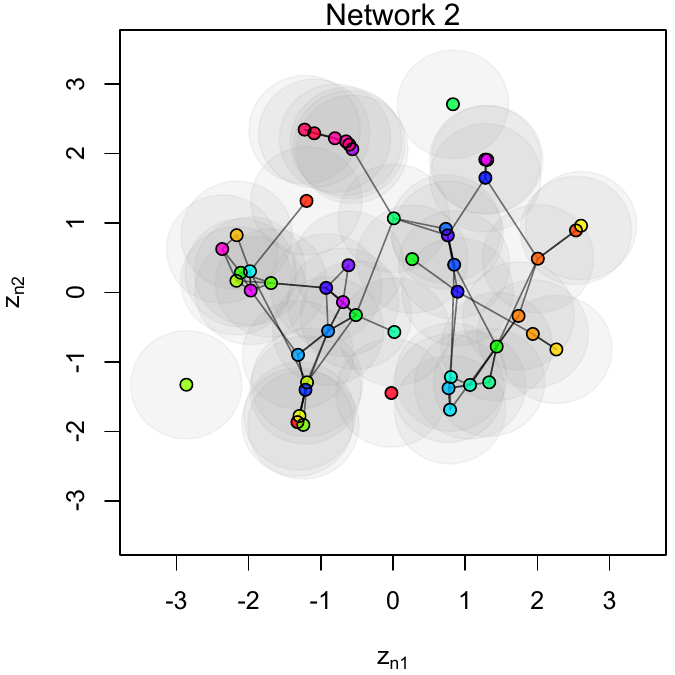}
\includegraphics[scale=.47]{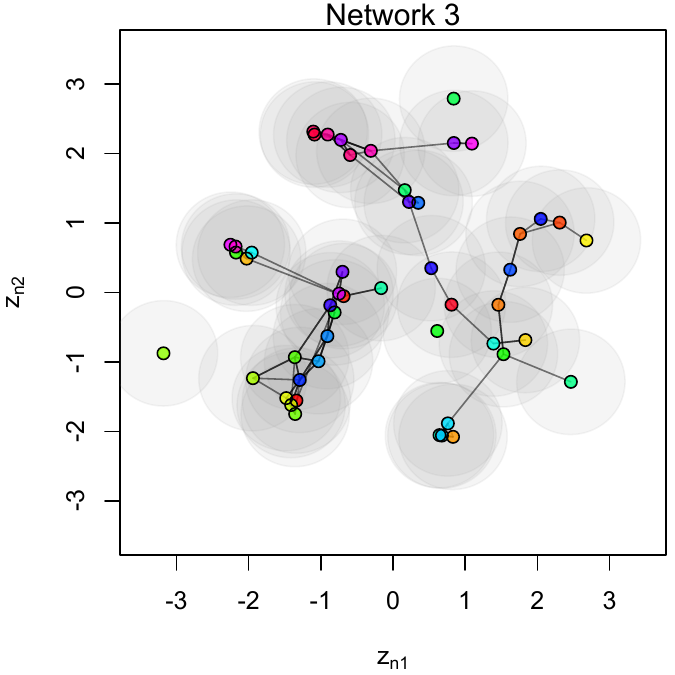}
\vspace{-.7cm}
\caption{Estimated latent positions under each model $\tilde{\mathbf{z}}_{i1},\tilde{\mathbf{z}}_{i2},\tilde{\mathbf{z}}_{i3}$ for the Girls Networks fitting the LSJM. The grey ellipses represent the $95 \%$ approximate credible intervals.
}\label{fig:girls.lsjm.lsm}
\end{center}
\end{figure}

\begin{figure}[!htp]
\begin{center}
\includegraphics[scale=0.5]{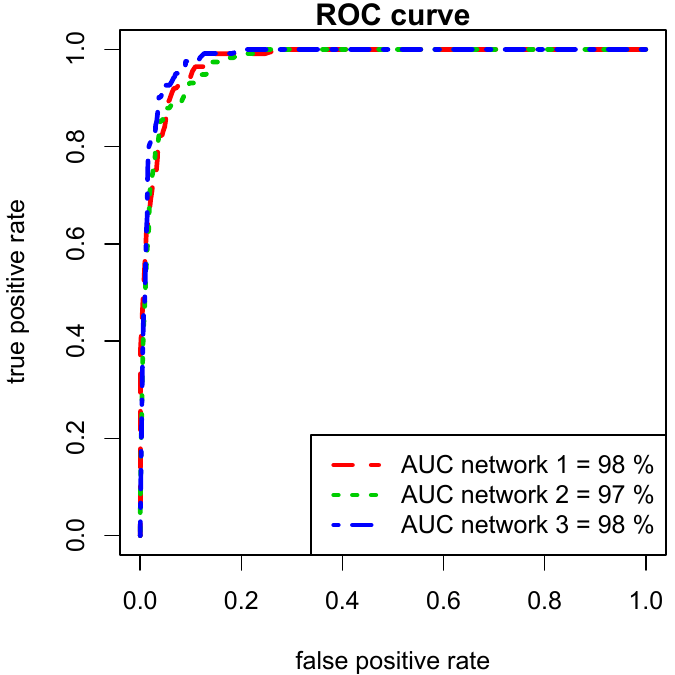}\hspace{.5cm}
\includegraphics[scale=0.5]{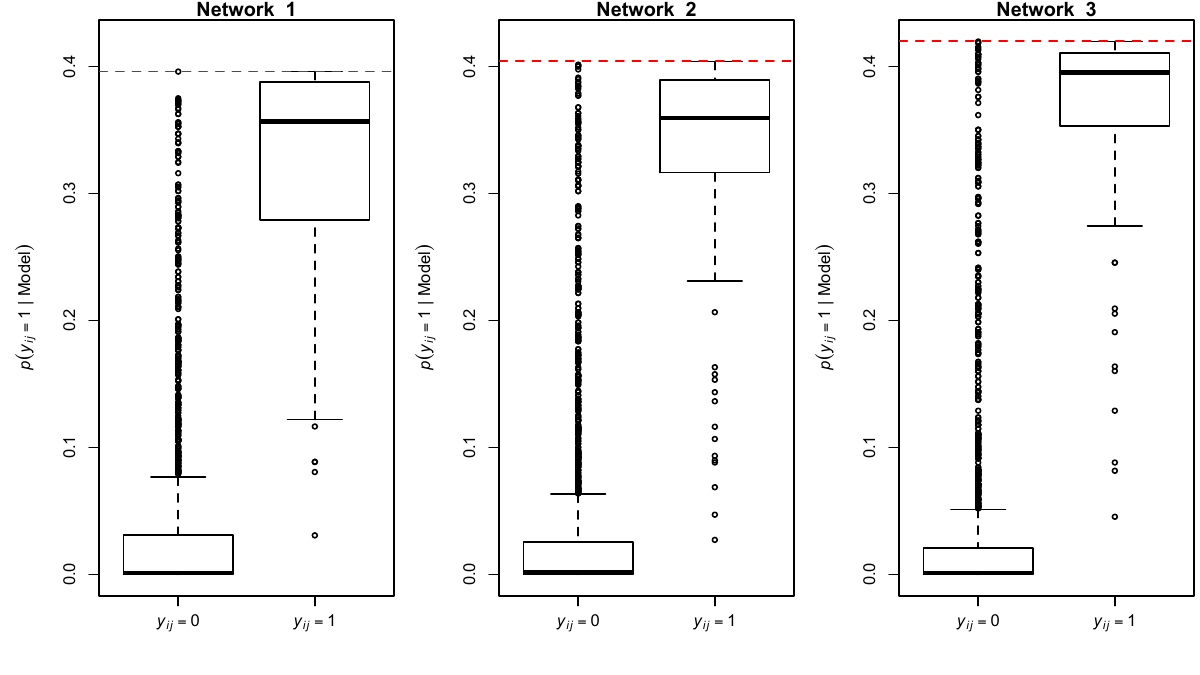}
\caption{ROC curves and boxplots of the estimated probabilities of a link for the true negatives and true positives obtained by fitting the LSJM to the Girls Networks.}\label{fig:girls.lsjm.roc}
\end{center}
\end{figure}

Figure~\ref{fig:girls.lsjm.lsm} shows the estimated positions $\ktbzi$ under each model in the latent space. These plots allow a direct comparison between the positions in the latent space $\ktbzi$ given by the LSJM and the ones obtained by fitting a single LSM for each network view (Figure~\ref{fig:girls.lsm}). It is interesting to observe that in the LSJM context the proximity between the disconnected components in network 3 depends on their previous relations.
The posterior probabilities of the model parameters $q(\alpha_1)=\mathcal{N}(-0.42,0.01)$, $q(\alpha_2)=\mathcal{N}(-0.39,0.01)$ and $q(\alpha_3)=\mathcal{N}(-0.32,0.01)$ are lower than the single LSM approaches (Section~\ref{sec.girls}), implying that a given distance $d$ between two nodes in the latent space corresponds to a higher probability of a link.
From the ROC curves, AUC and the boxplots (Figure~\ref{fig:girls.lsjm.roc}) of the estimated probabilities of a link, it appears clear that the LSJM fits the data quite well.
To evaluate the link prediction we applied a 10-fold cross validation setting $10\%$ of the links to be missing at each time point. The area under the ROC curves are 97, 96, 99\% fitting the LSJM and 89, 97, 98\% fitting the LSM, it shows that the estimates of the links are quite good, especially when applying the LSJM. 

Setting $p(\yijk=1|\ktbzi,\ktbzj,\txik)> \tau_k$ where $\tau_k$ is equal to the median probability of a link for the subgroup of the actual observed links in network $k$ and applying the LSJM we obtained a misclassification rate of $4\% $ for every network, whereas applying three single LSM we obtain a misclassification rate of $4\% $ for network 1, and $5\% $ for network 2 and 3.

\begin{figure}[!htp]
\begin{center}
\includegraphics[scale=.8]{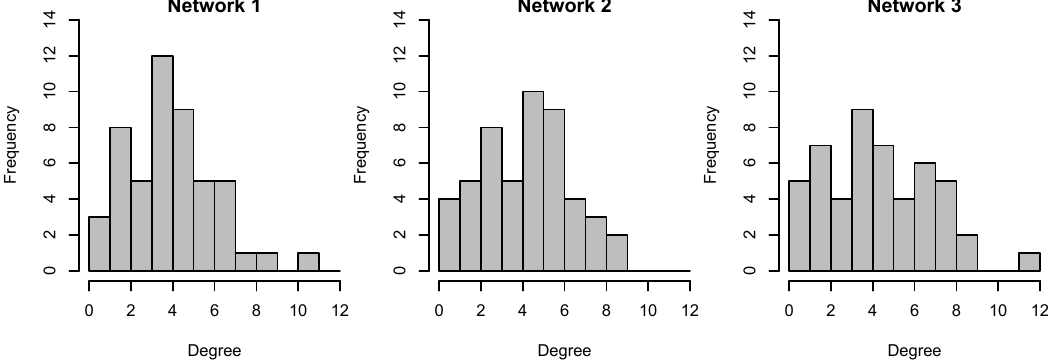}
\vspace{-.5cm}
\caption{Estimated Degree Distribution for the Girls Networks fitting the LSJM.}\label{fig:degrees50es}
\end{center}
\end{figure}

The estimated networks are formed of 109, 115 and 118 links respectively, and have density of 0.044, 0.047, 0.048. Their degree distributions are shown in Figure~\ref{fig:degrees50es}. Comparing these results with the true network statistics (Section~\ref{sec.girls}, and Figure~\ref{fig:degrees50}) it's possible to observe that the distributions are quite similar.

The LSJM allows one to also manage missing nodes, to do this we applied a 10-fold cross validation setting the $10\% $ of the nodes in each network to be missing. We obtained a misclassification rate of $9\% $ for all the three networks. As mentioned above in this case the LSM approach would be useless since it would locate the nodes only relying on the prior information.

\subsection{Saccharomyces Cerevisiae Protein-Protein Interactions}\label{sec.pro}

\begin{figure}[!htp]
\begin{center}
\includegraphics[scale=.8]{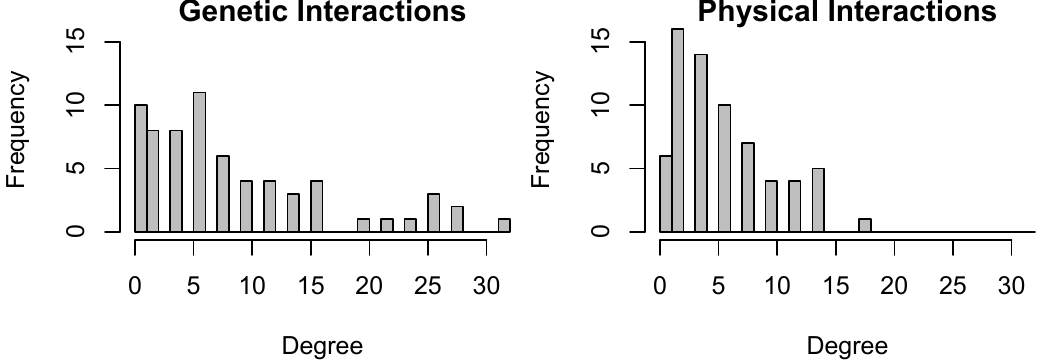}
\vspace{-.5cm}
\caption{Degree Distribution for the PPI datatset.}\label{fig:degreepro}
\end{center} 
\end{figure}

We analyse a dataset containing two undirected networks formed by genetic and physical protein-protein interactions (PPI) between 67 Saccharomyces cerevisiae proteins.
The genetic interactions network is formed of 294 links, and its density is 0.066, while the physical interactions network is formed of 190 links, and its density is 0.043. Their degree distributions are shown in Figure~\ref{fig:degreepro}. 
The complex relational structure of this dataset has led to implementation of models aiming at describing the functional relationships between the observations \citep{BKKI08,Troy03}.
A list of proteins included in this dataset is displayed in 
the supplementary material.
The dataset is available in the \texttt{lvm4net} package, and was downloaded from the Biological General Repository for Interaction Datasets (BioGRID) database\footnote{\url{http://thebiogrid.org/}} (\cite{Biogrid06}).  We refer to \cite{Biogrid06,Biogrid11} for a description of BioGRID, and for details regarding how the data were collected. 

\subsubsection{LSM}\label{sec.lsm.app.pro}
\begin{figure}[!htp]
\begin{center}
\includegraphics[scale=.47]{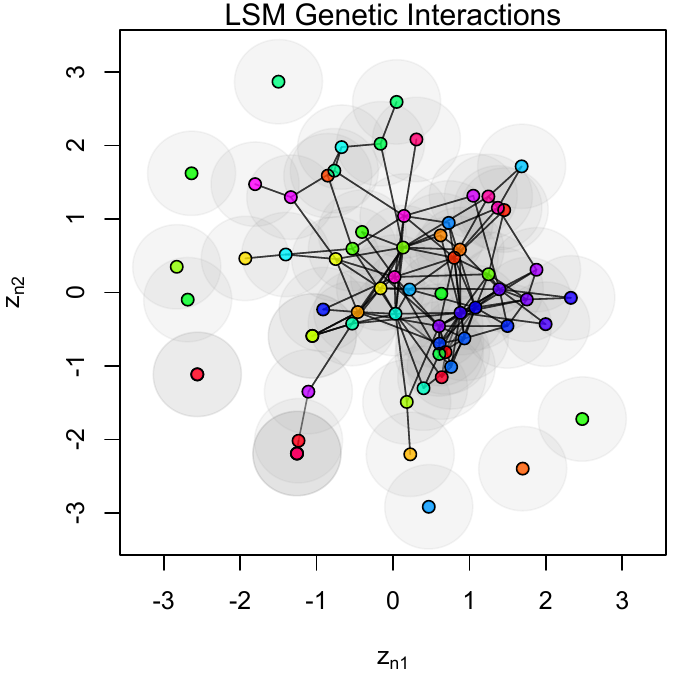}
\includegraphics[scale=.47]{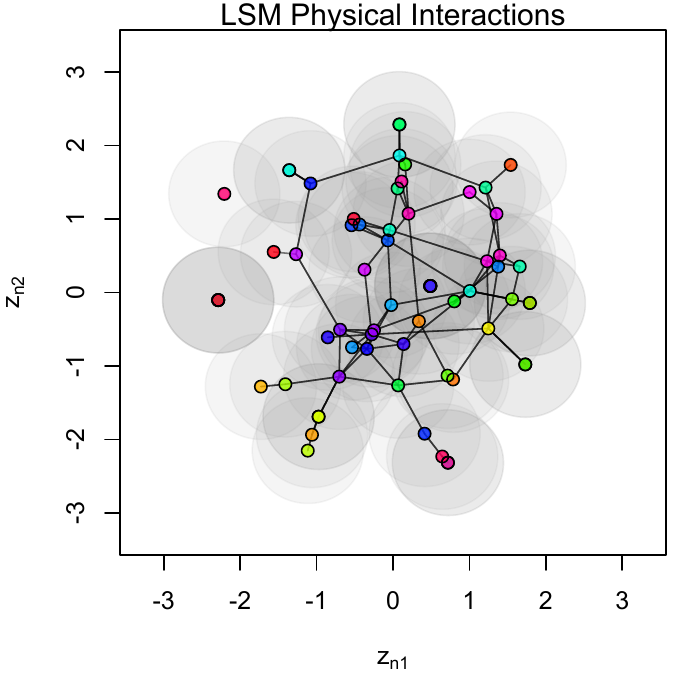}
\vspace{-.5cm}
\caption{Latent positions $\tbzi$ for the Genetic (left) and Physical (right) interaction data fitting the LSM. 
}\label{fig:pro.lsm}
\end{center}
\end{figure}

\begin{figure}[!htp]
\begin{center}
\includegraphics[scale=0.5]{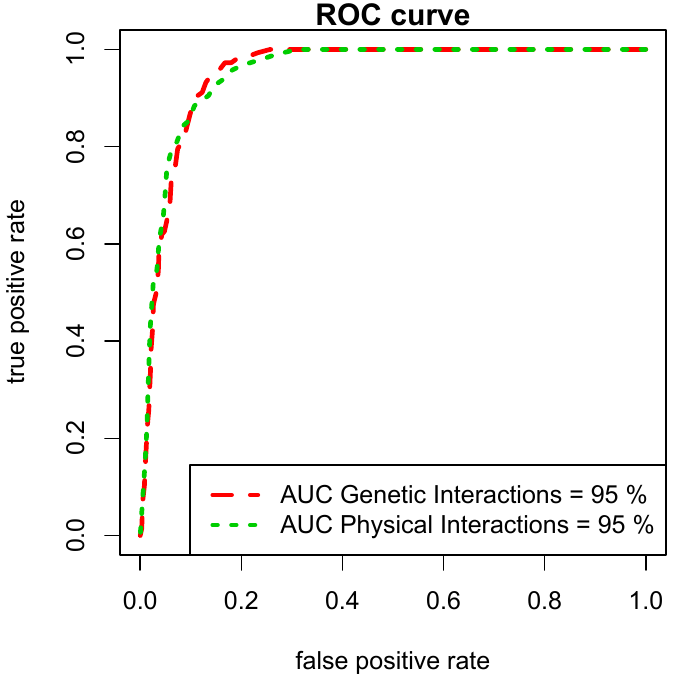}\hspace{2cm}
\includegraphics[scale=0.5]{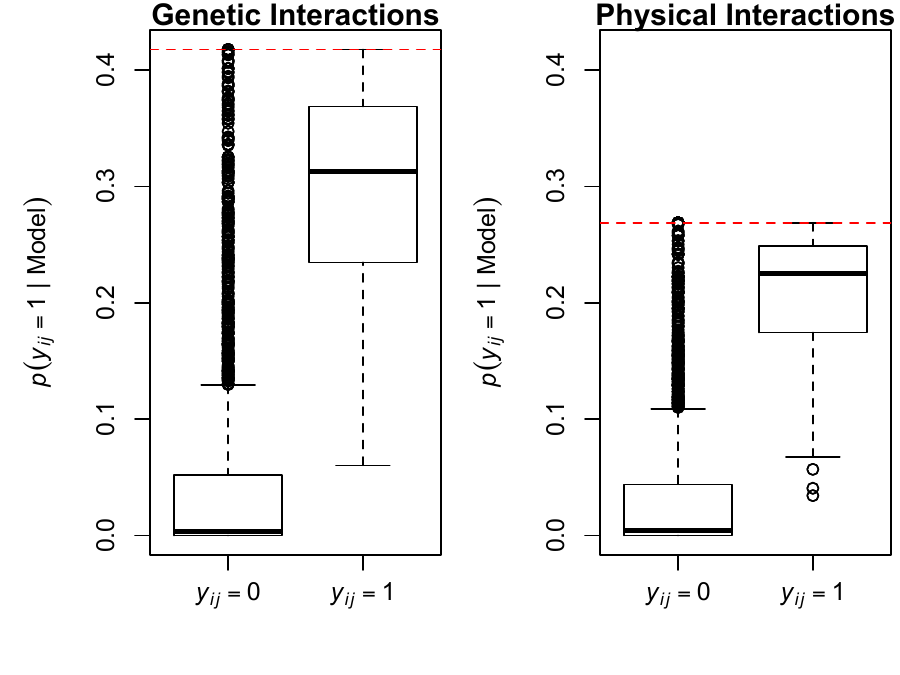}
\vspace{-.5cm}
\caption{ROC curves and boxplots of the estimated probabilities of a link for the true negatives and true positives obtained by fitting the LSM to the PPI data.}\label{fig:pro.lsm.roc}
\end{center}
\end{figure}

We fit the LSM to the PPI data working with the genetic and physical interaction networks separately.  We assume that $p(\alpha)=\mathcal{N}(0,2)$ and $p(\bzi)\overset{iid}{=}\mathcal{N}(\mathbf{0},\mathbf{I}_2)$.
The estimated latent positions $\tbzi$ (defined in Equation~\ref{eq:z.lsm}) are shown in Figure~\ref{fig:pro.lsm}.
The posterior distributions of the $\alpha$'s $q(\alpha_1)=\mathcal{N}(-0.332,0.003)$ and $q(\alpha_2)=\mathcal{N}(-1.001,0.005)$ indicate that network 1 regarding the genetic interactions is much more dense than network 2 which refers to the physical interactions.
The ROC curves, AUC and the boxplots (Figure~\ref{fig:pro.lsm.roc}) show that the proposed LSM fit the data quite well.

\subsubsection{LSJM}\label{sec.app.pro}
\begin{figure}[!htp]
\begin{center}
\includegraphics[scale=.47]{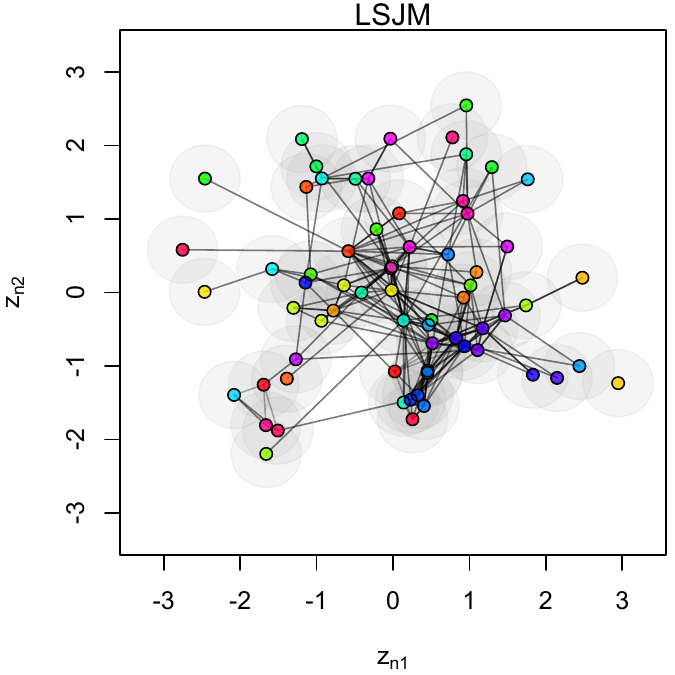}
\includegraphics[scale=.47]{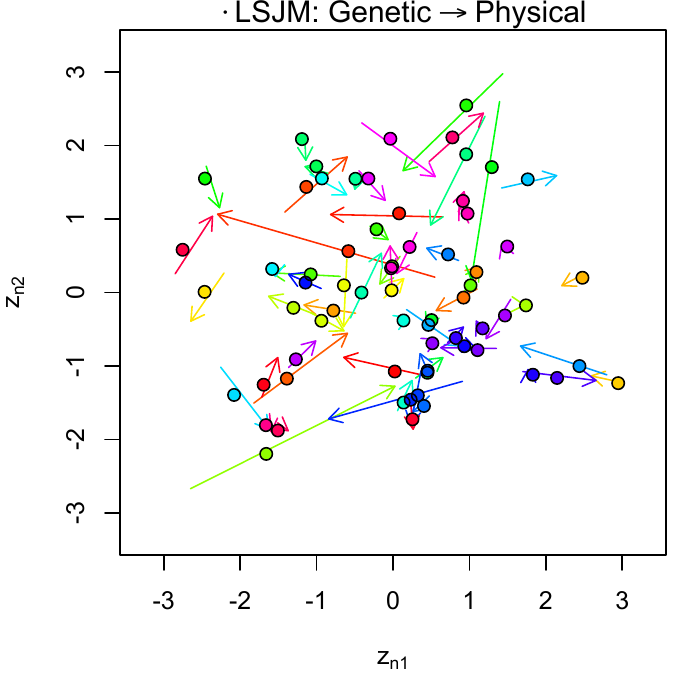}
\vspace{-.5cm}
\caption{On the left are shown the latent positions $\bbzi$ for the PPI networks fitting the LSJM. 
In the plot on the right the dots represent the overall positions $\bbzi$ and the arrows connect the estimated position under each model $\tilde{\mathbf{z}}_{i1},\tilde{\mathbf{z}}_{i2}$.}\label{fig:pro.lsjm}
\end{center}
\end{figure}

\begin{figure}[!htp]
\begin{center}
\includegraphics[scale=.47]{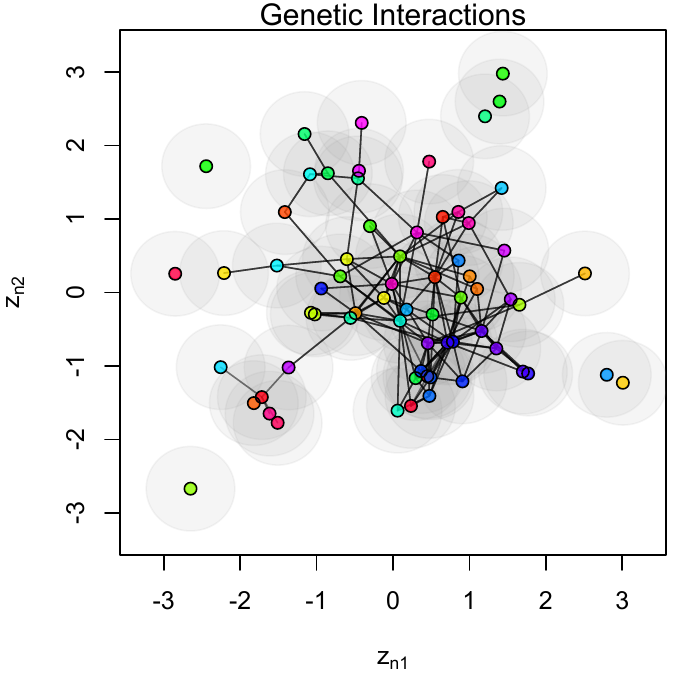}
\includegraphics[scale=.47]{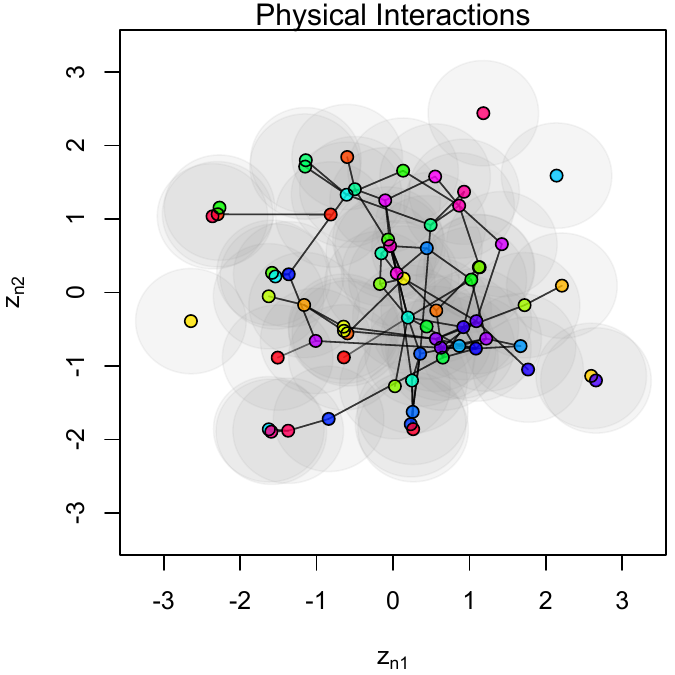}
\vspace{-.5cm}
\caption{Estimated latent positions under each model $\tilde{\mathbf{z}}_{i1},\tilde{\mathbf{z}}_{i2}$ for the PPI networks fitting the LSJM}\label{fig:pro.lsjm.lsm}
\end{center}
\end{figure}

\begin{figure}[!htp]
\begin{center}
\includegraphics[scale=0.5]{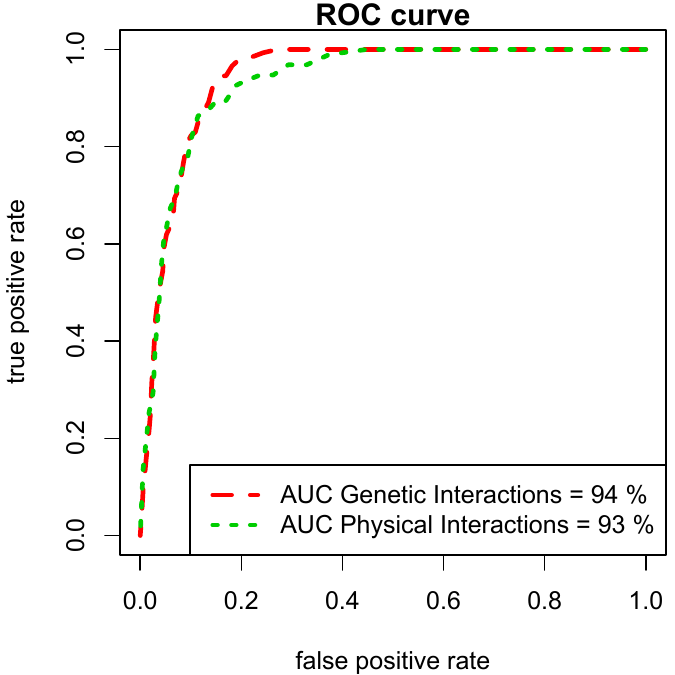}\hspace{2cm}
\includegraphics[scale=0.5]{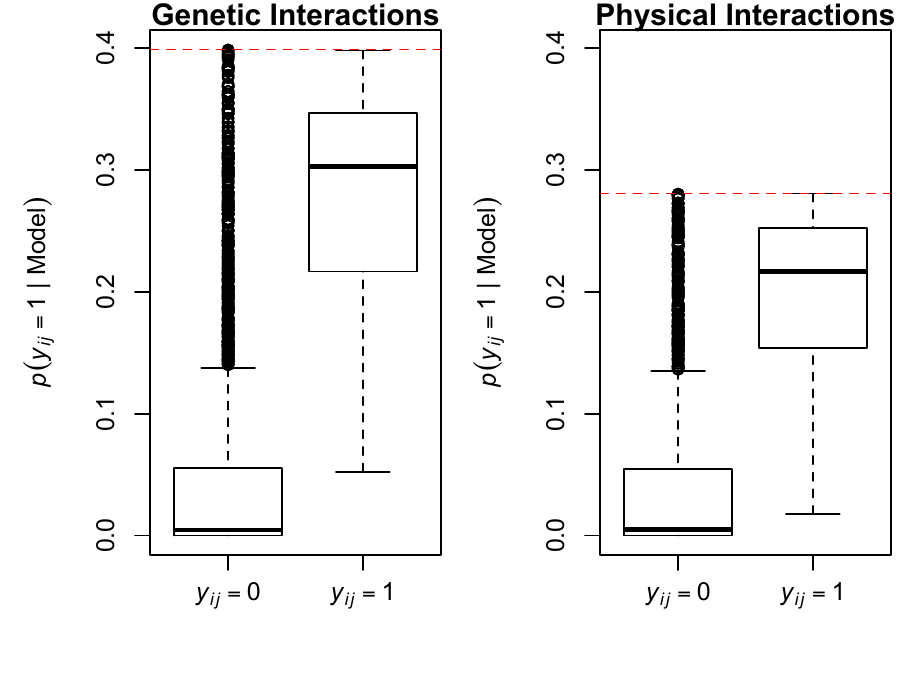}
\vspace{-.5cm}
\caption{ROC curves of the estimated probabilities of a link for the true negatives and true positives obtained by fitting the LSJM to the PPI Networks}\label{fig:pro.lsjm.roc}
\end{center}
\end{figure}

We apply the LSJM to the PPI dataset.
Figure~\ref{fig:pro.lsjm} shows the estimated overall latent positions $\bbzi$ (defined in Equation~\ref{eq:zb.lsjm}) and in the plot on the right each arrow starts from the positions $\tbziI$ for the genetic interaction dataset and points to the latent positions $\tilde{\mathbf{z}}_{i2}$ for the physical interaction dataset  (defined in Equation~\ref{eq:z.lsjm}).
Figure~\ref{fig:pro.lsjm.lsm} shows the estimated position under each model $\ktbzi$ in the latent space. It is possible to compare these results with Figure~\ref{fig:pro.lsm} in which we have fitted two single LSM to the data. 
The ROC curve, AUC and the boxplots in Figure~\ref{fig:pro.lsjm.roc} show that the LSJM fits the data quite well. The posterior distributions for the $\alpha$'s are $q(\alpha_1)=\mathcal{N}(-0.410,0.003)$ $q(\alpha_2)=\mathcal{N}(-0.940,0.005)$.

We applied a 10-fold cross validation to evaluate the prediction of missing links. The area under the ROC curve is 94\% for both the genetic and the physical interaction networks fitting the LSJM, fitting the LSM the AUC values are 66\% for the genetic interaction network and 97\% for the physical interaction network. The results of the LSJM show a much better fit in terms of estimates of missing links for the genetic interaction network compared to the ones obtained by the single LSM approach. Setting the threshold to be equal to the median probability of a link for the subgroup of the actual observed links we obtained similar misclassification rates for the missing links in both LSJM and single LSM approach: $9\% $ for the genetic interaction network, and $6\% $ for the physical interaction network. 

\begin{figure}[!htp]
\begin{center}
\includegraphics[scale=.8]{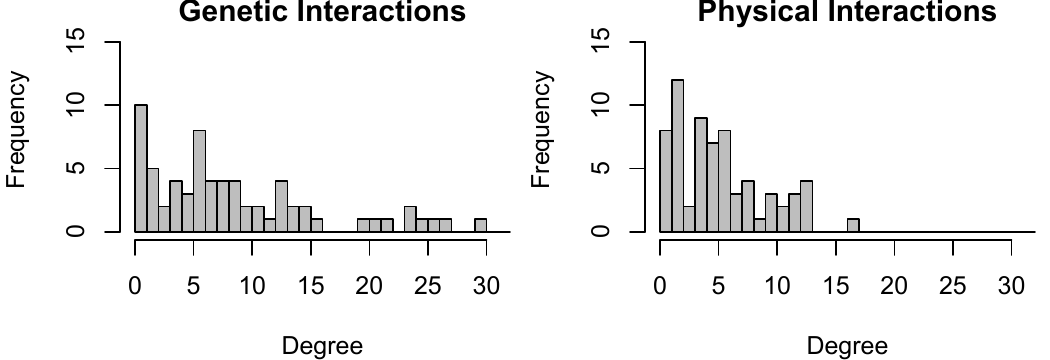}
\vspace{-.5cm}
\caption{Estimated Degree Distribution for the PPI networks fitting the LSJM.}\label{fig:degreeproes}
\end{center}
\end{figure}

The estimated genetic interactions network is formed of 297 links, and its density is 0.067, while the estimated physical interactions network is formed of 185 links, and its density is 0.042. Their degree distributions are shown in Figure~\ref{fig:degreeproes}. Comparing these results with the true network statistics (Section~\ref{sec.pro}) it's possible to observe that the simulated degree distributions (Figure~\ref{fig:degreeproes}) are smoother then the original one (Figure~\ref{fig:degreepro}), but overall they are quite similar. 

Applying a 10-fold cross validation for missing nodes using the LSJM we obtained a misclassification rate of $24\% $ for the genetic interactions dataset and $20\%$ for the physical interaction network.

\section{Conclusions}\label{sec.concl}

A lot of network data require the introduction of novel models able to describe their complex connectivity structure. On the other hand new inferential methods are needed to carry out estimation efficiently.
In this paper, we proposed a latent variable model (LSJM) for multiple network views that extends the latent space model proposed by \cite{HRH02}, allowing the information given by different relations on the same nodes to be summarized in the same latent space. 
The use of the variational approach to compute the model parameters allows us to apply the latent space joint model to larger networks (of the order of thousands of nodes). 
A comparison between the variational method and  MCMC for the single view network latent space model is given in \cite{SalterTownshend2009}. 
An alternative variational algorithm that can be used in this context could be derived from the methods outlined in \cite{OpAr2009}. This model allows the position of each observation in a latent space to be found based on all the available information in the datasets. 
Further information like the latent positions in each network view separately are obtained updating the single-network estimates given the overall positions. This information allows effective visualization and prediction of the data. The examples presented show how the LSJM facilitates the interpretation of the different positions of the network nodes in the latent space according to longitudinal measurements in the excerpt of 50 girls from `Teenage Friends and Lifestyle Study' example  and multiple relations in the Saccharomyces Cerevisiae genetic and physical protein-protein interactions dataset.
All the methods presented in this paper are included in the \texttt{lvm4net} package for \texttt{R} \citep{lvm4net}. 

Future work may lead to an extension of the model allowing cluster formation by assuming that the latent positions come from a Gaussian mixture model fitting each network using a Latent Position Cluster Model \citep{HRT07}. Or a novel modelling approach which takes explicitly into account the sequential feature as in the dynamic network analysis \citep{Sar05,Hoff11,WH11}.

\section*{Acknowledgements}

The authors wish to acknowledge the anonymous reviewers for helpful comments. This work was supported by Science Foundation funded Clique Research Cluster (08/SRC/I1407) and Insight Research Centre (SFI/12/RC/2289). 


\newpage
\setcounter{section}{0}
\setcounter{figure}{0}
\setcounter{table}{0}
\setcounter{page}{1}

\begin{center}
{\LARGE \bf Supplementary Material for\\
 ``Joint Modelling of Multiple Network Views''}\\[.3cm]

        {\large Isabella Gollini$ \qquad \qquad \qquad $  Thomas Brendan Murphy}

\end{center}

\section{Empirical comparison of squared Euclidean model and non-squared Euclidean model}

We perform an empirical analysis in order to compare the visualization and prediction properties of the squared Euclidean distance model to the non-squared Euclidean model.

We use the following \texttt{R} packages: \texttt{latentnet} \citep{Krivitsky2008, latentnet}, \texttt{VBLPCM} \citep{SalterTownshend2009, vblpcm} and \texttt{lvm4net} \citep{lvm4net}. Their main features are shown in Table~\ref{tab:pack}. [Timings can be considerably improved in the \texttt{lvm4net} package by converting some functions into \texttt{C}.]

The latent space model is defined as:
\begin{equation*}
p(\bY|\bZ,\alpha)=\prod_{i\neq j}^N \frac{\exp(\alpha-d_{ij})^{\yij}}{1+\exp(\alpha-d_{ij})}
\end{equation*}

The Euclidean distance (ED in Table~\ref{tab:pack}) is defined as: $d_{ij} = |\bzi-\bzj|$; the Squared Euclidean distance (SED in Table~\ref{tab:pack}) is defined as: $d_{ij} = |\bzi-\bzj|^2$.

\begin{table}[ht]
\centering
\caption{Comparison of the main features of the packages for latent space modeling}\label{tab:pack}
\begin{tabular}{|l|cc|cc|}
  \hline
 & \multicolumn{2}{c|}{\textbf{Model}} &\multicolumn{2}{c|}{\textbf{Inference Method}}\\
& ED & SED & MCMC & Variational \\ 
  \hline
\texttt{latentnet} & \cmark & \xmark & \cmark & \xmark \\ 
\texttt{VBLPCM} & \cmark & \xmark & \xmark & \cmark \\ 
\texttt{lvm4net} & \xmark & \cmark & \xmark & \cmark \\
   \hline
\end{tabular}
\end{table}

\subsection{Connected component of the genetic PPI network}
\begin{figure}[!htp]
\begin{center}
\includegraphics[scale=.9]{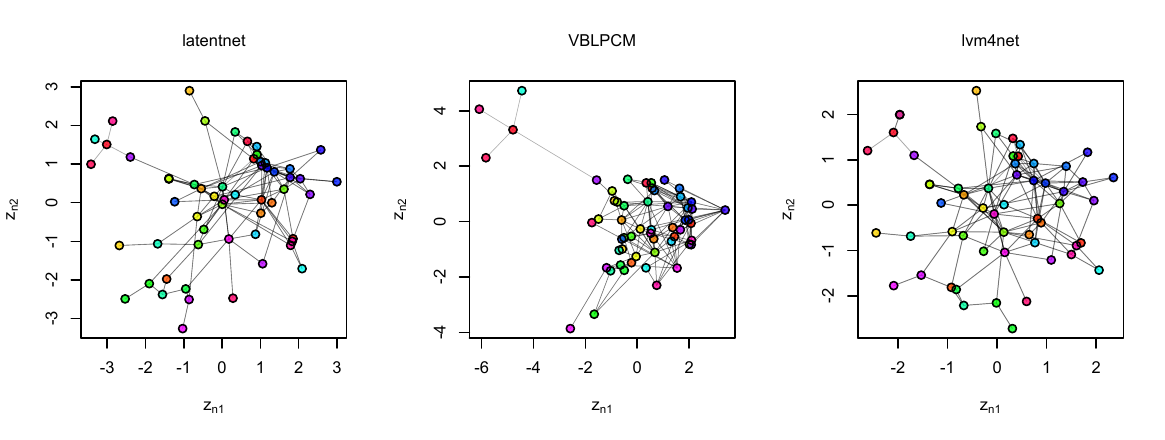}
\vspace{-.5cm}
\caption{Latent positions for the most connected component of the genetic PPI network}\label{fig:compare}
\end{center} 
\end{figure}

\begin{figure}[!htp]
\begin{center}
\includegraphics[scale=.9]{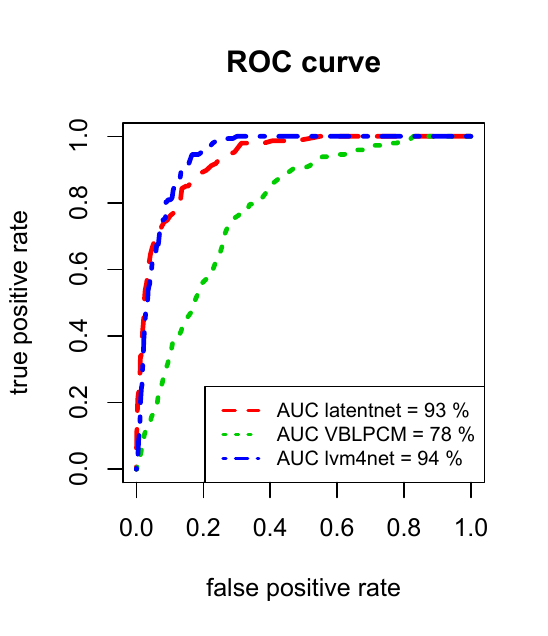}
\vspace{-.5cm}
\caption{ROC curves of the estimated probabilities of a link for the true negatives and true positives obtained for the most connected component of the genetic PPI network}\label{fig:compRoc}
\end{center}
\end{figure}

\begin{figure}[!htp]
\begin{center}
\includegraphics[scale=.9]{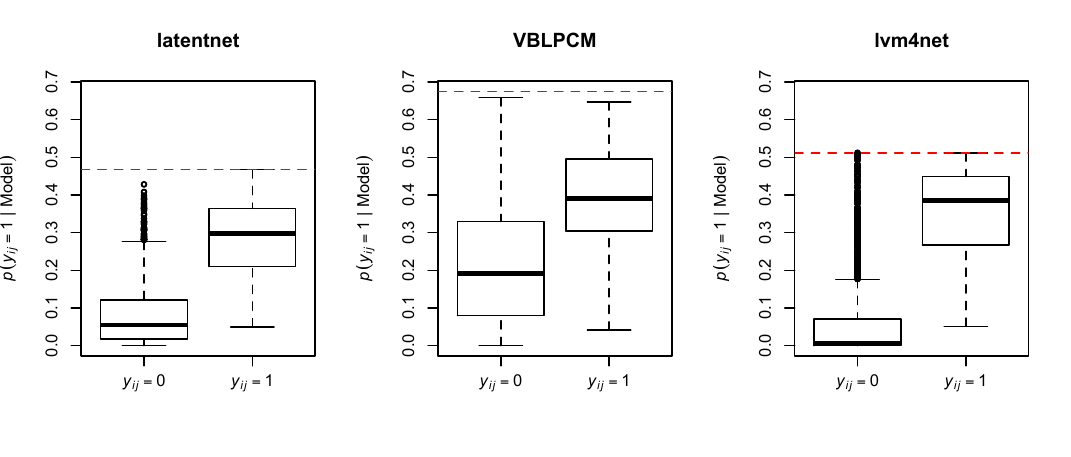}
\vspace{-.5cm}
\caption{Boxplot of the estimated probabilities of a link for the true negatives and true positives obtained for the most connected component of the genetic PPI network}\label{fig:compBox}
\end{center}
\end{figure}

In order to compare the results obtained by the three methods we use the connected component of the genetic PPI network described in Section 5.3 of the paper. This component consists of 57 nodes and 294 edges. Table~\ref{tab:compare} shows the CPU times employed to fit the models. It is evident that the \texttt{lvm4net} package is much faster than the \texttt{latentnet} package.
In Figure~\ref{fig:compare} it is possible to see that the positions obtained using \texttt{latentnet} and \texttt{lvm4net} are extremely similar, and both approaches fit the data very well (Figures~\ref{fig:compRoc} and \ref{fig:compBox}) while \texttt{VBLPCM} provides poorer results.

\begin{table}[ht]
\centering
\caption{Timings in seconds to fit the models to the most connected component of the genetic PPI network}\label{tab:compare}
\begin{tabular}{l|c}
  \hline
& Time in sec. \\ 
  \hline
MCMC - Euclidean distance (\texttt{latentnet}) & 73.17 \\ 
Variational - Euclidean distance (\texttt{VBLPCM}) & 7.90 \\ 
Variational - Squared Euclidean distance (\texttt{lvm4net}) & 1.68 \\ 
   \hline
\end{tabular}
\end{table}

\subsection{Datasets used in the main paper}

Since the \texttt{VBLPCM} package is only working with connected components, in Table~\ref{tab:compareAUC} and ~\ref{tab:compareTime} we only compare the results obtained from the network datasets described in the paper using \texttt{latentnet} and \texttt{lvm4net}. Both \texttt{latentnet} and \texttt{lvm4net} perform well in terms of AUC scores.

\begin{table}[!htb]
    \begin{minipage}{.5\linewidth}
      \centering
      \caption{AUC}\label{tab:compareAUC}
\begin{tabular}{l|cc}
  \hline
& \texttt{latentnet} & \texttt{lvm4net}\\ 
  \hline
   $Y_1$ & 0.99  & 0.98  \\ 
   $Y_2$ & 0.99  & 0.98  \\ 
   $Y_3$ & 0.99  & 0.99 \\ 
   \hline
 $Y_{gen}$ & 0.96 & 0.95 \\ 
  $Y_{phy}$ & 0.97 & 0.95 \\ 
  \hline
\end{tabular}
    \end{minipage}%
    \begin{minipage}{.5\linewidth}
      \centering
\caption{Timings in Seconds}\label{tab:compareTime}
\begin{tabular}{l|cc}
  \hline
& \texttt{latentnet} & \texttt{lvm4net}\\ 
  \hline
     $Y_1$ & 62.98 & 1.02 \\ 
      $Y_2$ & 65.94 & 1.33  \\ 
   $Y_3$ & 67.65 & 1.21 \\ 
   \hline
$Y_{gen}$ & 97.21 & 1.80\\ 
  $Y_{phy}$ & 104.73 & 1.48\\ 
  \hline
\end{tabular}
    \end{minipage} 
\end{table}

\newpage
\section{Saccharomyces Cerevisiae Protein-Protein Interactions Data}\label{sec.pro}

\begin{figure}[!htp]
\begin{center}
\includegraphics[scale=.8]{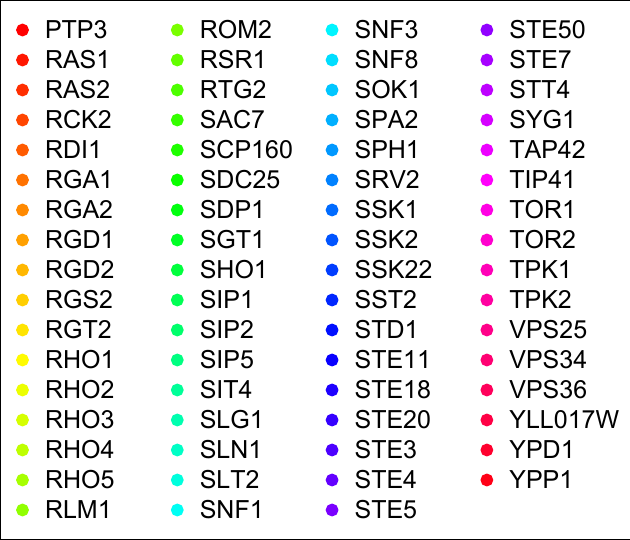}
\vspace{-.5cm}
\caption{Saccharomyces cerevisiae proteins}\label{fig:pro}
\end{center}
\end{figure}

We analyse a dataset containing two undirected networks consisting of genetic and physical protein-protein interactions between 67 Saccharomyces cerevisiae proteins.
A list of proteins included in this dataset is displayed in Figure~\ref{fig:pro}.
The data were downloaded from the Biological General Repository for Interaction Datasets (BioGRID) database\footnote{\url{http://thebiogrid.org/}} (\cite{Biogrid06}).  We refer to \cite{Biogrid06,Biogrid11} for a description of BioGRID, and for details regarding how the data were collected.

\section{Estimation procedures for the LSM parameters} \label{app.par}
\subsection{E-Step LSM}
\begin{itemize}
\item[1] Estimate $\tbzi$

\begin{equation*}
\begin{split}
\mathrm{KL}&\leq \mathrm{Const}_{\tbzi}+\tbzi^T\tbzi \left(\frac{1}{2\sigma^2}+\sum_{j \neq i}(y_{ji}+y_{ij})\right) -2\tbzi^T\left(\sum_{i \neq n}(y_{ji}+y_{ij})\tbzj\right) \\
&\quad +2 \sum_{j \neq i}\log \left(1+ \frac{\exp\left(\txi+\dfrac{1}{2}\tpsi^2\right)}{ \det(\bI+4\tSigma)^{\frac{1}{2}}}\exp\left(-(\tbzi-\tbzj)^T(\bI+4\tSigma)^{-1}(\tbzi-\tbzj)\right)\right)
\end{split}
\end{equation*}
Second-order Taylor series expansion approximation of
\begin{equation*}
f(\tbzi)=\sum_{j \neq i}\log \left(1+ \frac{\exp\left(\txi+\dfrac{1}{2}\tpsi^2\right)}{ \det(\bI+4\tSigma)^{\frac{1}{2}}}\exp\left(-(\tbzi-\tbzj)^T(\bI+4\tSigma)^{-1}(\tbzi-\tbzj)\right)\right)
\end{equation*}
Therefore,
\begin{equation*}
f(\tbzi)\approx f(\tbzio)+(\tbzi-\tbzio)^TG(\tbzio)+\dfrac{1}{2}(\tbzi-\tbzio)^T H(\tbzio) (\tbzi-\tbzio)
\end{equation*}
Let find the gradient $G$ and the Hessian matrix $H$ of $f$ evaluated at $\tbzi=\tbzio$.
\begin{equation*}
G(\tbzio)=-2(\bI+4\tSigma)^{-1}\sum_{j \neq n}(\tbzio-\tbzj)\left[1+\frac{ \det(\bI+4\tSigma)^{\frac{1}{2}}}{\exp\left(\txi+\dfrac{1}{2}\tpsi^2\right)}\exp\left((\tbzio-\tbzj)^T(\bI+4\tSigma)^{-1}(\tbzio-\tbzj)\right)\right]^{-1}
\end{equation*}
\begin{equation*}
\begin{split}
H(\tbzio)=-2(\bI+4\tSigma)^{-1}\sum_{j \neq n}&\left[1+\frac{ \det(\bI+4\tSigma)^{\frac{1}{2}}}{\exp\left(\txi+\dfrac{1}{2}\tpsi^2\right)}\exp\left((\tbzio-\tbzj)^T(\bI+4\tSigma)^{-1}(\tbzio-\tbzj)\right)\right]^{-1} \cdot \\
\cdot & \left[\bI-
\frac{2(\tbzio-\tbzj)(\tbzio-\tbzj)^T(\bI+4\tSigma)^{-1}}{1+ \frac{\exp\left(\txi+\dfrac{1}{2}\tpsi^2\right)}{ \det(\bI+4\tSigma)^{\frac{1}{2}}}\exp\left(-(\tbzio-\tbzj)^T(\bI+4\tSigma)^{-1}(\tbzio-\tbzj)\right)}
\right]
\end{split}
\end{equation*}

Therefore,
\begin{equation*}
\mathrm{KL}\approx \tbzi^T \left[\left(\frac{1}{2\sigma^2}+\sum_{j \neq i}(y_{ji}+y_{ij})\right)\bI + H(\tbzio)\right] \tbzi -2\tbzi^T\left[\sum_{j \neq i}(y_{ji}+y_{ij})\tbzj-G(\tbzio)+H(\tbzio) \tbzio \right] +\mathrm{Const}_{\tbzi}
\end{equation*}
$\dfrac{\partial \mathrm{KL}}{\partial \tbzi}=0$

\begin{equation*}
\tbzi=\left[ \left(\frac{1}{2\sigma^2}+\sum_{j \neq i}(y_{ji}+y_{ij})\right)\bI + H(\tbzio) \right]^{-1}\left[\sum_{j \neq i}(y_{ji}+y_{ij})\tbzj-G(\tbzio)+H(\tbzio) \tbzio \right]
\end{equation*}

\item[2]  Estimate $\tSigma$

\begin{equation*}
\begin{split}
\mathrm{KL}&\leq \mathrm{Const}_{\tSigma}+\mathrm{tr}(\tSigma)\left( \frac{N}{2\sigma^2}+2\sum_{i=1}^N\sum_{j\neq i}\yij \right) - \frac{N}{2} \log(\mathrm{det}(\tSigma)) \\
&\quad +\sum_{i=1}^N\sum_{j\neq i}\log\left(1+ \frac{\exp\left(\txi+\dfrac{1}{2}\tpsi^2\right)}{ \det(\bI+4\tSigma)^{\frac{1}{2}}}\exp\left(-(\tbzi-\tbzj)^T(\bI+4\tSigma)^{-1}(\tbzi-\tbzj)\right)\right)
\end{split}
\end{equation*}

First-order Taylor series expansion approximation of: 

\begin{equation*}
f(\tSigma)=\sum_{i=1}^N\sum_{j\neq i}\log\left(1+ \frac{\exp\left(\txi+\dfrac{1}{2}\tpsi^2\right)}{ \det(\bI+4\tSigma_0)^{\frac{1}{2}}}\exp\left(-(\tbzi-\tbzj)^T(\bI+4\tSigma_0)^{-1}(\tbzi-\tbzj)\right)\right)
\end{equation*}
\begin{equation*}
f(\tSigma)\approx f(\tSigma_0)+J(\tSigma_0)(\tSigma-\tSigma_0)
\end{equation*}
where $J$ is the Jacobian matrix of $f$ evaluated at $\tSigma=\tSigma_0$.

\begin{equation*}
\begin{split}
J(\tSigma_0)&=4(\bI+4\tSigma_0)^{-1}\sum_{i=1}^N\sum_{j\neq i}\left((\tbzi-\tbzj)(\tbzi-\tbzj)^T(\bI+4\tSigma_0)^{-1}-\frac{1}{2}\bI\right)  \cdot\\
& \quad \cdot \left[1+\frac{ \det(\bI+4\tSigma_0)^{\frac{1}{2}}}{\exp\left(\txi+\dfrac{1}{2}\tpsi^2\right)}\exp\left((\tbzi-\tbzj)^T(\bI+4\tSigma_0)^{-1}(\tbzi-\tbzj)\right)\right]^{-1}
\end{split}
\end{equation*}

Therefore,

\begin{equation*}
\mathrm{KL}\approx \mathrm{tr}(\tSigma)\left( \frac{N}{2\sigma^2}+2\sum_{i=1}^N\sum_{j\neq i}\yij \right) - \frac{N}{2} \log(\mathrm{det}(\tSigma)) + J(\tSigma_0)\tSigma  +\mathrm{Const}_{\tSigma}
\end{equation*}

$\dfrac{\partial \mathrm{KL}}{\partial \tSigma}=0$

\begin{equation*}
\tSigma=\frac{N}{2}\left[ \left(\frac{N}{2\sigma^2}+2\sum_{i=1}^N\sum_{j\neq i}\yij\right)\bI +J(\tSigma_0) \right]^{-1}
\end{equation*}

\end{itemize}

\subsection{M-Step LSM}
\begin{itemize}
\item[1] Estimate $\txi$

\begin{equation*}
\mathrm{KL}\leq \frac{\txi^2}{2\psi^2} -\txi\left( \frac{\xi}{\psi^2}+\sum_{i=1}^N\sum_{j\neq i} \yij\right) + \sum_{i=1}^N\sum_{j\neq i}\log \left(1+\exp(\txi)A_{i,j}\right)+\mathrm{Const}_{\txi}
\end{equation*}
where $A_{i,j}=\exp\left(\dfrac{1}{2}\tpsi^2\right) \det(\bI+4\tSigma)^{-\frac{1}{2}}\exp\left(-(\tbzi-\tbzj)^T(\bI+4\tSigma)^{-1}(\tbzi-\tbzj)\right)$.\\

Second-order Taylor series expansion of:
\begin{equation*}
f(\txio)=\sum_{i=1}^N\sum_{j\neq i}\log \left(1+\exp(\txio)A_{i,j}\right)
\end{equation*}
evaluated at $\txi=\txio$

\begin{equation*}
f(\txi)\approx f(\txio)+f'(\txio) (\txi-\txio)+\dfrac{1}{2}f''(\txio)(\txi-\txio)^2
\end{equation*}

where
\begin{equation*}
f'(\txio)=\sum_{i=1}^N\sum_{j\neq i}\left(1+\exp(-\txio)A_{i,j}^{-1}\right)^{-1}
\end{equation*}
\begin{equation*}
f''(\txio)=\sum_{i=1}^N\sum_{j\neq i}\left(1+\exp(-\txio)A_{ij}^{-1}\right)^{-1}\left(1+\exp(\txio)A_{i,j}\right)^{-1}
\end{equation*}
Therefore,

\begin{equation*}
\mathrm{KL}\leq \frac{1}{2}\txi^2\left( \frac{1}{\psi^2}+f''(\txio)\right) -\txi\left( \frac{\xi}{\psi^2}+\sum_{i=1}^N\sum_{j\neq i} \yij-f'(\txio)+\txio f''(\txio)\right) +\mathrm{Const}_{\txi}
\end{equation*}

$\dfrac{\partial \mathrm{KL}}{\partial \txi}=0$

\begin{equation*}
\txi=\frac{\xi+\psi^2(\sum_{i=1}^N\sum_{j\neq i} \yij-f'(\txio)+\txio f''(\txio))}{1+\psi^2 f''(\txio)}
\end{equation*}

\item[2] Estimate $\tpsi^2$

\begin{equation*}
\mathrm{KL}\leq \frac{\tpsi^2}{2\psi^2}-\frac{1}{2} \log(\tpsi^2) + \sum_{i=1}^N\sum_{j\neq i}\log \left(1+\exp\left(\frac{1}{2}\tpsi^2\right)A_{i,j}\right)+\mathrm{Const}_{\tpsi^2}
\end{equation*}
where $A_{i,j}=\exp(\txi) \det(\bI+4\tSigma)^{-\frac{1}{2}}\exp\left(-(\tbzi-\tbzj)^T(\bI+4\tSigma)^{-1}(\tbzi-\tbzj)\right)$

First-order Taylor series expansion of:
\begin{equation*}
f(\tpsi^2)=\sum_{i=1}^N\sum_{j\neq i}\log \left(1+\exp\left(\frac{1}{2}\tpsi_{0}^2\right)A_{i,j}\right)
\end{equation*}
evaluated at $\tpsi^2=\tpsi_{0}^2$

\begin{equation*}
f(\tpsi^2)\approx f(\tpsi_{0}^2)+f'(\tpsi_{0}^2) (\tpsi^2-\tpsi_{0}^2)
\end{equation*}

where
\begin{equation*}
f'(\tpsi_{0}^2)=\sum_{i=1}^N\sum_{j\neq i}\frac{1}{2}\left(1+\exp\left(-\frac{1}{2}\tpsi_{0}^2\right)A_{i,j}^{-1}\right)^{-1}
\end{equation*}

Therefore,

\begin{equation*}
\mathrm{KL}\approx \tpsi^2\left(\frac{1}{2\psi^2} + f'(\tpsi^2_0) \right)- \frac{1}{2}\log(\tpsi^2) +\mathrm{Const}_{\tpsi^2}
\end{equation*}

$\dfrac{\partial \mathrm{KL}}{\partial \tpsi^2}=0$

\begin{equation*}
\tpsi^2=\left(\frac{1}{\psi^2} +2 f'(\tpsi^2_0) \right)^{-1}
\end{equation*}
\end{itemize}

\end{document}